\pdfoutput=1% This requests pdfLaTeX typeset for arXiv
\documentclass[aps,prb,reprint,superscriptaddress]{revtex4-1}

\usepackage{graphicx}% Include figure files
\usepackage{epstopdf}
\usepackage{color}

\draft % marks overfull lines with a black rule on the right

\begin{document}

%\preprint{}

\title{Vanishing fine structure splittings in telecom wavelength quantum dots grown on (111)A surfaces by droplet epitaxy} 

\author{Xiangming~Liu}
%\thanks{}
\affiliation{National Institute for Materials Science, 1-1 Namiki, Tsukuba 305-0044, Japan}
\author{Neul~Ha}
\affiliation{National Institute for Materials Science, 1-1 Namiki, Tsukuba 305-0044, Japan}
\affiliation{Graduate School of Engineering, Kyushu University, NIMS, Tsukuba 305-0044, Japan}
\author{Hideaki~Nakajima}
\affiliation{Research Institute for Electronic Science, Hokkaido University, Sapporo 001-0021, Japan}
\author{Takaaki~Mano}
\affiliation{National Institute for Materials Science, 1-1 Namiki, Tsukuba 305-0044, Japan}
\author{Takashi~Kuroda}
\email[]{kuroda.takashi@nims.go.jp}
\affiliation{National Institute for Materials Science, 1-1 Namiki, Tsukuba 305-0044, Japan}
\affiliation{Graduate School of Engineering, Kyushu University, NIMS, Tsukuba 305-0044, Japan}
\author{Bernhard~Urbaszek}
\affiliation{Universit\'{e} de Toulouse, INSA-CNRS-UPS, LPCNO, 135 Av.~Rangueil, 31077 Toulouse, France}
\author{Hidekazu~Kumano}
\author{Ikuo~Suemune}
\affiliation{Research Institute for Electronic Science, Hokkaido University, Sapporo 001-0021, Japan}
\author{Yoshiki~Sakuma}
\author{Kazuaki~Sakoda}
\affiliation{National Institute for Materials Science, 1-1 Namiki, Tsukuba 305-0044, Japan}

\date{\today}

\begin{abstract} 
The emission cascade of a single quantum dot is a promising source of entangled photons. A prerequisite for this source is the use of a symmetric dot analogous to an atom in a vacuum, but the simultaneous achievement of %high 
structural symmetry and emission in a telecom band poses a challenge. Here we report the growth and characterization of highly symmetric InAs/InAlAs quantum dots self-assembled on $C_{3v}$ symmetric InP(111)A. The broad emission spectra cover the O ($\lambda \sim 1.3$~$\mu$m), C ($\lambda \sim 1.55$~$\mu$mm), and L ($\lambda \sim 1.6$~$\mu$m) telecom bands. The distribution of the fine-structure splittings %reveals an average value of 25~$\mu$eV, which 
is considerably smaller than those reported in previous works on dots at similar wavelengths. The presence of dots with degenerate exciton lines is further confirmed by the optical orientation technique. Thus, our dot systems are expected to serve as efficient entangled photon emitters for long-distance fiber-based quantum key distribution.

%%
%%
%The emission cascade of a single quantum dot is a promising source of entangled photons. A prerequisite for this source is the use of a symmetric dot analogous to an atom in vacuum, but simultaneous achievement of high structural symmetry and emission in telecom band poses a challenge. Here we report the growth and characterization of highly symmetric InAs/InAlAs quantum dots self-assembled on $C_{3v}$ symmetric InP(111)A. 
%Making use of droplet epitaxy, we create laterally symmetric InAs/InAlAs quantum dots self-assembled on $C_{3v}$ symmetric InP(111)A. 
%The broad emission spectra cover the telecom O ($\lambda \sim 1.3$~$\mu$m), C ($\lambda \sim 1.55$~$\mu$m), and L ($\lambda \sim 1.6$~$\mu$m) bands. Distribution of fine-structure splittings reveals an average value of 25~$\mu$eV, which is considerably smaller than previous works on dots at similar wavelengths. The optical orientation technique enables to confirm the presence of dots with smaller fine-structure splitting than the expected natural width of 0.5~$\mu$eV. (1.4~ns in lifetime). The presence of dots with degenerate exciton lines is further confirmed by optical orientation technique. Thus, our dot systems are expected to serve as efficient entangled photon emitters for long-distance fiber-based quantum key distribution.
\end{abstract}


\maketitle %\maketitle must follow title, authors, abstract and \pacs
%%%%%%%%%
%%%%%%%%%
%\noindent\textbf{\textit{Introduction.}} 
Semiconductor quantum dots (QD) are expected to play a central role in quantum information networks. A noteworthy device based on dots is the solid-state single photon source, which ensures absolute security in quantum key distribution (QKD) \cite{Zeillinger_RMP99,*Gisin_RMP02}. Since QDs can confine charged carriers in nanometer-sized regions, recombination enables single photons to appear on demand, i.e., synchronously with a master clock shared in networks \cite{Benson_PRL00}. QKD over a 50 km commercial fiber has already been demonstrated with QD photon sources, which emitted at a wavelength of 1.5~$\mu$m \cite{Takemoto_APEX10}. The transmission distance in that work was limited purely by the absorption loss of silicate fibers. Exceeding this fundamental limit requires the development of quantum link protocols, which exploit the nonlocality inherent in quantum theory. An efficient source of entangled photon pairs is a key element in the realization of such protocols, examples of which include quantum teleportation \cite{Bennett_PRL93} and entanglement swapping \cite{Zukowski_PRL93,*Bose_PRA98,*Pan_PRL98}. 

The generation of entangled photons with semiconductor QDs is directly linked to the singlet configuration of two excitons (X), which form a biexciton (XX). Eventually, two photons associated with the XX-X cascade show polarization correlations independent of the choice of measurement basis, yielding quantum entanglement in the polarization state. However, a common class of QDs exhibits considerable fine-structure splittings (FSS) \cite{Gammon_PRL96,Bayer_PRL99,Kulakovskii_PRL99,Seguin2005,Marco_PRB08}, which exclude entanglement in emitted photons \cite{Santori_PRB02}. Numerous attempts have been made to suppress FSS and recover the symmetry of QDs grown on conventional (001) oriented substrates \cite{Lagbein_PRB04,*Young_PRB05,Akopian_PRL06,Stevenson_Nat06,*Pooley_PRAppl14,Seidl_APL06,*Gerardot_APL07,Muller_PRL08,Dousse_Nat10,Ghali_NatComm12,*Pooley_APL13,Trotta_PRL12,*Trotta_arxiv14}. However, from a practical point of view, the reproducible growth of symmetric dots with (at least) near-zero FSS is highly desirable. 

A noteworthy strategy for achieving such high QD symmetry is the application of $C_{3v}$ symmetric (111) surfaces to the growth substrate, as was predicted theoretically \cite{Sing_PRL09,*Schliwa_PRB09}. Although QD growth in the Stranski-Krastanov (SK) mode is prohibited along the [111] axis, the use of patterned substrates \cite{Sugiyama_APL95,*Hartmann_APL98,*Mereni_APL09} and droplet epitaxy \cite{Stock_APL10,Mano2010,*Jo_CGD12} makes it possible to grow QDs on (111) substrates. Hence, a great reduction in FSS was observed in these QDs \cite{Mano2010,*Jo_CGD12,Karlsson_PRB10,*Mereni_PRB10}, which led to the demonstration of entangled photon emission in pyramidal QDs on patterned (111)B substrates \cite{Juska2013}, and the filtering-free violation of Bell's inequality for droplet epitaxial GaAs/AlGaAs dots on GaAs(111)A \cite{Kuroda2013}. Note that all these studies dealt with visible wavelength photons. Material challenges have meant that the development of QD sources in telecom bands has achieved less progress. The simultaneous realization of small FSS and telecom-band emission is a great challenge. 

%A remarkable strategy for achieving such high QD symmetry is the application of $C_{3v}$ symmetric (111) surfaces to the growth substrate, as was predicted theoretically \cite{Sing_PRL09,Schliwa_PRB09}. Though QD growth with the Stranski-Krastanov (SK) mode is prohibited along [111] axis, the use of patterned substrates \cite{Sugiyama_APL95,Hartmann_APL98,*Mereni_APL09} and droplet epitaxy \cite{Stock_APL10,Mano2010,*Jo_CGD12} enables to grow QDs on (111) substrates. Hence, great reduction in FSS was observed in these QDs \cite{Mano2010,*Jo_CGD12,Karlsson_PRB10,*Mereni_PRB10}, which led to the demonstration of entangled photon emission in pyramidal QDs on patterned (111)B substrates \cite{Juska2013}, and filtering-free violation of Bell's inequality for droplet epitaxial GaAs/AlGaAs dots on GaAs(111)A \cite{Kuroda2013}. Note that all these works dealt with visible wavelength photons. Due to material challenges, development of QD sources in telecom band has been less progressed. Simultaneous realization of small FSS and telecom-band emission poses a great challenge. 

In this work, we report on the growth and characterization of telecom-band InAs quantum dots on (111)A substrates. For this purpose, we focus on the use of droplet epitaxy, which is not strain driven, thereby enabling us to choose a variety of materials and surface orientations. %
Though most previous works on droplet epitaxy dealt with lattice-matched systems, the versatility of this technique makes it possible to create QDs on lattice-mismatched systems targeting telecom-band emission. %
The successful growth of InAs/InAlAs QDs on InP(111)A has recently been demonstrated \cite{Ha_APL14}. Here we use newly created QDs with a lower surface density, which allows a systematic study of FSS and the symmetry characteristics of dots. The measured FSS reveals an average value of 25~$\mu$eV, which is considerably smaller than those in previous studies %undertaken 
on SK grown dots at similar wavelengths \cite{Cade_PRB06,Chauvin2006,Sapienza2013}. Moreover, the presence of QDs with nearly perfect exciton degeneracy is confirmed using the optical orientation technique. Thus our source is expected to serve as a promising candidate for highly efficient entangled photon sources, which do not require the use of serious temporal gating to improve the degree of quantum correlation \cite{Ward_natcomm14}. 

%In this work, we report on the growth and characterization of telecom-band InAs quantum dots on (111)A substrates. For this purpose, we focus on the use of droplet epitaxy, which is not strain driven, thereby capable of choosing a variety of materials and surface orientations. Successful growth of InAs/InAlAs QDs on InP(111)A has recently been demonstrated \cite{Ha_APL14}. Here we use newly-created QDs with lower surface density, which allows systematic study on the FSS and symmetry characteristics of dots. The measured FSS reveals an average value of 25~$\mu$eV, which is considerably smaller than those in previous works at similar wavelengths \cite{Cade_PRB06,Chauvin2006,Sapienza2013}. Moreover, the presence of QDs with nearly-perfect exciton degeneracy is confirmed using optical orientation technique. Thus our source is expected to serve as a promising candidate for highly efficient entangled photon sources, which do not require the use of serious temporal gating to improve the degree of quantum correlations \cite{Ward_natcomm14}.

\begin{figure}
\includegraphics[width=7cm]{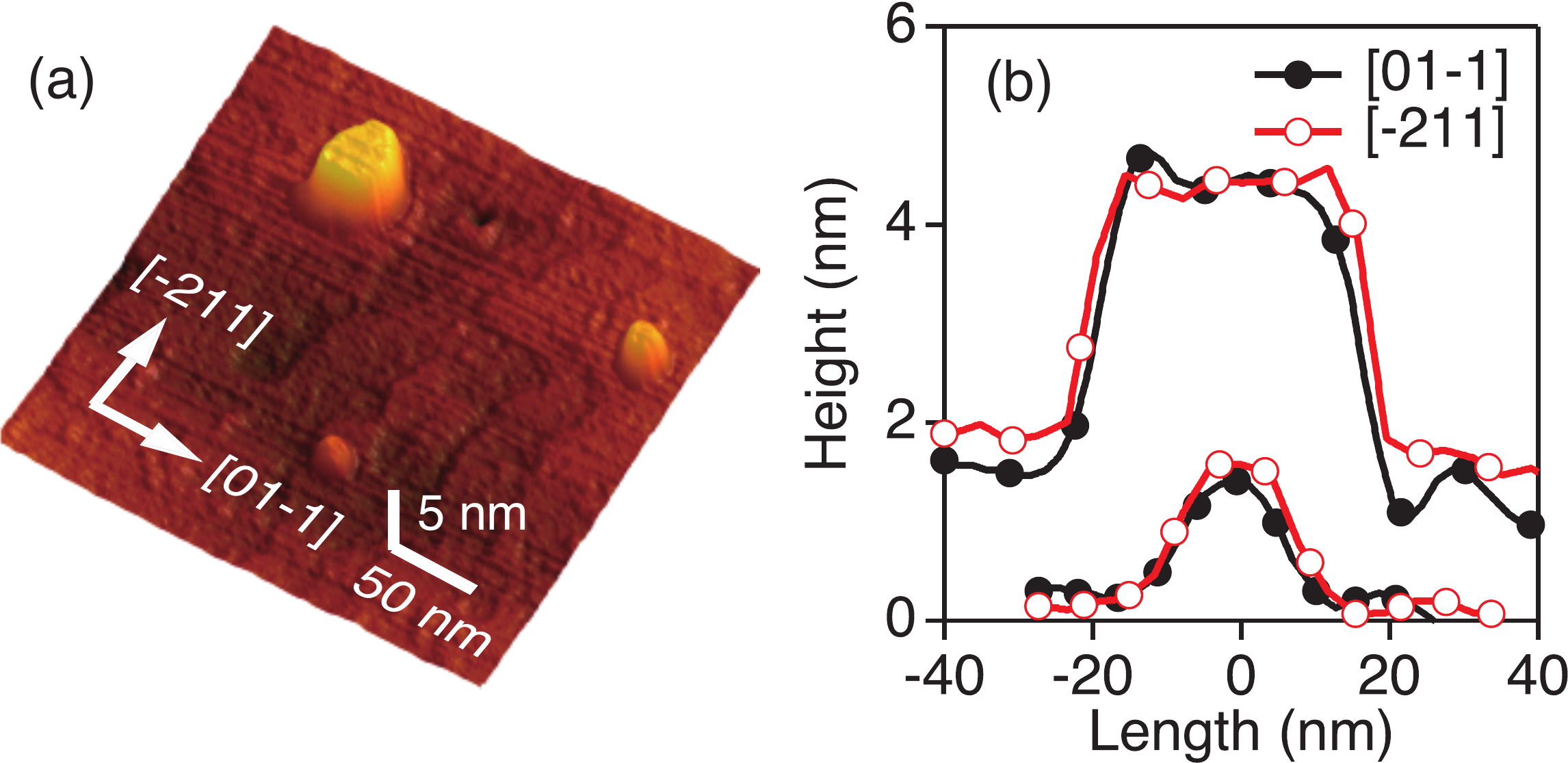}%
\caption{\label{afm}(color online) (a) AFM image of the sample surface of InAs dots on In$_{0.52}$Al$_{0.48}$As/InP(111)A. (b) Cross sections along [01-1] and [-210] for a relatively large dot (top) and a small dot (bottom). }
\end{figure}

%%%%%%%
%%%%%%%
%\vspace{1em}
%\noindent\textbf{\textit{Experimental procedure.}}
The sample investigated in this study was grown by droplet epitaxy using a conventional molecular beam epitaxy apparatus \cite{Koguchi1991,*Mano_Nanotechnology09,Sallen_PRL11,*Sallen_NatComm14}. After growing a 150-nm-thick In$_{0.52}$Al$_{0.48}$As barrier layer on an InP(111)A substrate at 470$^{\circ}$C, we cooled the substrate to 320$^{\circ}$C and supplied 0.4 monolayers of indium, which led to the formation of indium droplets. Next, we supplied an As$_4$ flux of 3$\times$10$^{-5}$~Torr to crystallize the indium droplets into InAs dots at 270$^{\circ}$C. The sample was then annealed at 370$^{\circ}$C and capped with another In$_{0.52}$Al$_{0.48}$As barrier layer with a thickness of 75 nm. %Finally, the thermal annealing process was done at 470~$^{\circ}$C to improve optical properties. 

For atomic force microscopy (AFM) analysis, an additional QD layer was grown on the top of the sample. Figure~\ref{afm}(a) shows a three-dimensional view of the surface, which reveals the presence of disk-like dots with $3.0\,(\pm 1.0)$ nm in hight and $38\,(\pm 10)$ nm in diameter. The dot density is as low as $3.2 \times 10^9$ cm$^{-2}$, which makes it possible to isolate single dots using conventional microphotoluminescence techniques. Figure~\ref{afm}(b) shows the AFM cross sections for two example QDs. Cross sections obtained along the orthogonal in-plane directions, [01-1] and [-211], are almost identical, which supports the high lateral symmetry in the dot shape without any elongations. This symmetric characteristic is a consequence of QD growth on (111) substrates. 

For the PL measurement we used a continuous wave laser emitting at a wavelength of 705 nm for excitation above the barrier band gap. The laser light was focused on the sample using a near-infrared microscope objective with a numerical aperture of 0.65. To reduce the spot size, a hemispherical solid immersion lens with a refractive index of two was positioned on the sample. Spontaneously emitted photons were collected with the same objective, and then fed into a 50-cm focal length polychromator equipped with an %thermo-electrically cooled 
InGaAs array detector. The spectrometer had a resolution of 55~$\mu$eV (0.08 nm) with a full width at half maximum (FWHM) at a wavelength of 1.3~$\mu$m. The linearly polarized PL spectra were recorded as a function of the polarization angle. %using a rotatable half-wave plate and a calcite polarizer. 
With a Gaussian fit to the emission lines, we were able to determine the spectral peak shift (and the absolute value of FSS) with a resolution as high as 4~$\mu$eV. All the experiments were performed at 10~K. 

\begin{figure}
\includegraphics[width=8cm]{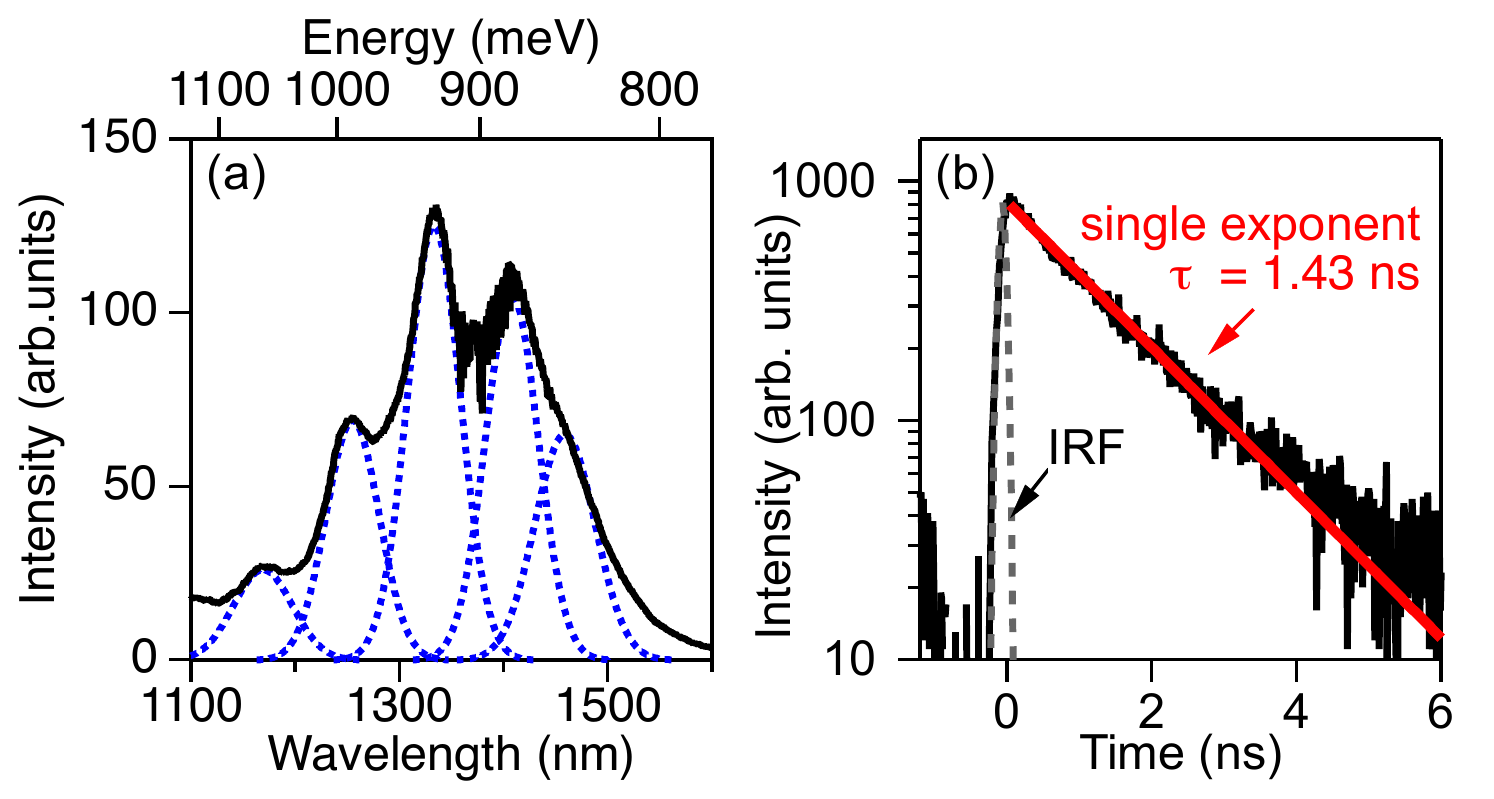}%
\caption{\label{ensemble}(color online) (a) PL spectrum of the ensemble of dots. The blue broken lines show the results of multiple-peak fit with assuming Gaussian broadening for each peak. (c) Time-resolved PL signals after ps pulsed excitation. The gray broken line shows the instrumental response function (IRF). The red line shows a single exponential fit to the decay data with a lifetime of 1.43~ns. }
\end{figure}

%%%%%%%%%
%%%%%%%%%
%\vspace{1em}
%\noindent\textit{\textbf{Ensemble spectrum and dynamics.}} 
Figure~\ref{ensemble}(a) shows the low-temperature PL spectrum of the dot ensembles. The PL spectrum spreads in a 1.1 to 1.6~$\mu$m wavelength range, which covers the O ($\lambda \sim1310$ nm), C ($\lambda \sim1550$ nm), and L ($\lambda \sim1600$ nm) telecom bands. The spectrum consists of several split peaks, among which high-yield emissions are centered at $\sim$930~meV (1333 nm) with an FWHM of 40 meV. The appearance of multiple peaks can be attributed to the different families of QDs with heights varying in monolayer steps \cite{Mano2010}. The AFM analysis suggests that the QDs have a flat shape with a height that ranges from 2 to 6 monolayers, which is consistent with the observed spectral profile.

%Figure~\ref{ensemble}(a) shows the low-temperature PL spectrum of the dot ensembles. The PL spectrum spreads in the range between 1.1~$\mu$m and 1.6~$\mu$m, which covers the telecom O- ($\lambda \sim1310$ nm), C- ($\lambda \sim1550$ nm), and L bands ($\lambda \sim1600$ nm). It consists of several split peaks, among which high-yield emissions are centered at $\sim$930 meV (1333 nm) with an FWHM of 40 meV. The appearance of multiple peaks can be attributed to the different families of QDs with heights varying by a monolayer step \cite{Mano2010}. The AFM analysis suggests that the QDs have a flat shape with a height which ranges from 2 to 6 monolayeres, being consistent with the observed spectral profile.

Figure~\ref{ensemble}(b) shows the PL decay signals of the dot ensembles at wavelengths around 1.3~$\mu$m after pulsed excitation. For this measurement, we used a mode-locked Ti sapphire laser for excitation ($\lambda \sim 785$~nm) and a superconducting single photon detector (SSPD) for detection. The decay curve reveals a single exponent with a decay time of 1.43~ns, which agrees with the theoretical decay time of spontaneous emission on the assumption that the QDs have the same dipole moment as the bulk value. The similar decay times have been confirmed in telecom-wavelength QDs grown for different substrate orientations \cite{Zinoni_APL06,*Takemoto_JAP07,*Miska_APL08}. Thus, the observed PL decay is likely governed by intrinsic carrier recombination and free of any non-radiative process, as a consequence of the high crystalline quality of dots. The homogeneous linewidth, which gives the maximum limit of FSS for entangled photon emission, is thus $\sim0.5$~$\mu$eV for our dots.

\begin{figure}
\includegraphics[width=8.5cm]{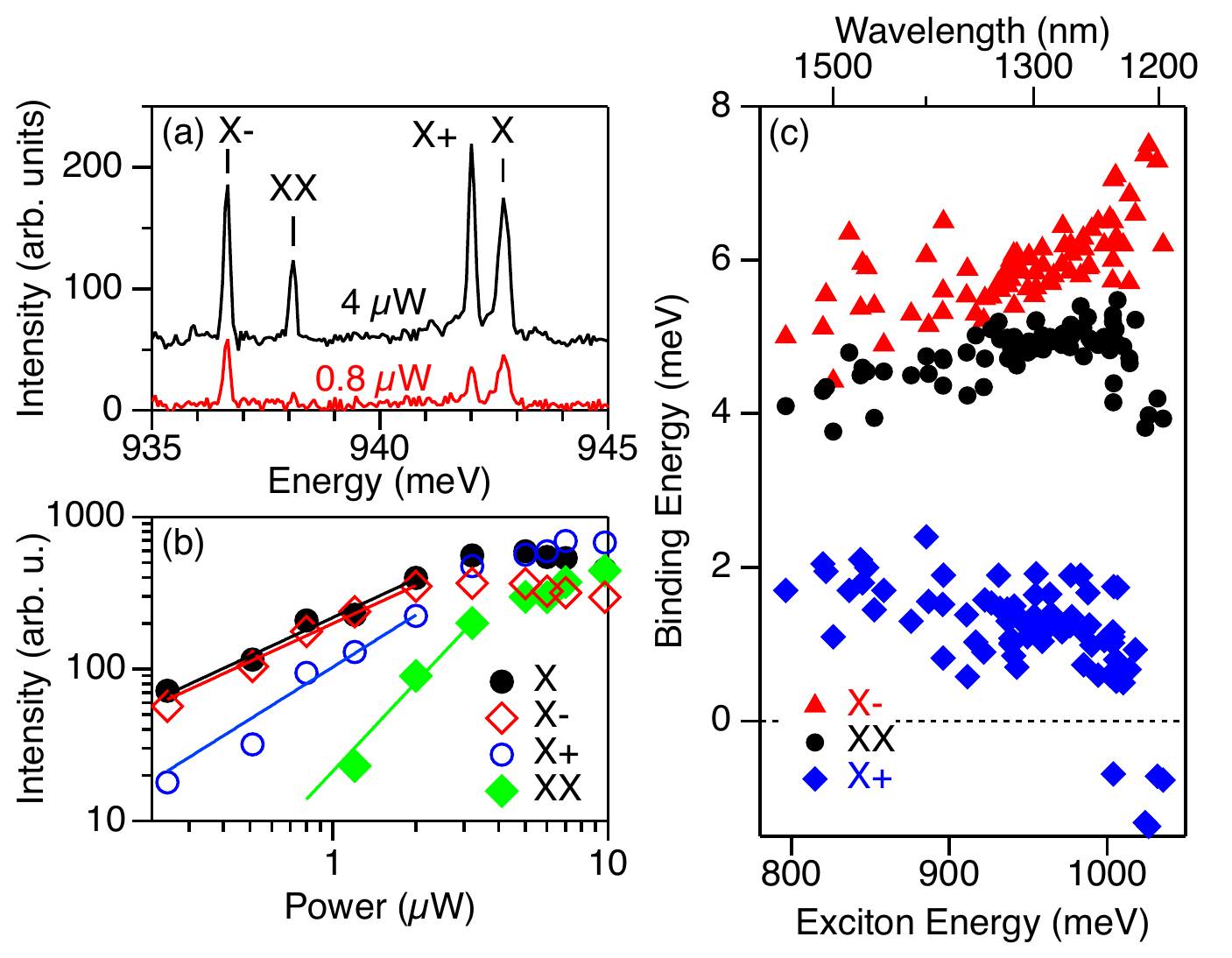}%
\caption{\label{multiX}(color online) (a) Typical PL spectra of a single QD with an excitaion power of 4~$\mu$W and 0.8~$\mu$W. (b) PL intensity as a function of excitation power for each exciton complex. The solid lines show power lows, $I\propto P^{\alpha}$, where $\alpha = 0.8,\,0.8,\,1.1$, and $1.8$ for X, X$^-$, X$^+$, and XX, respectively. (c) Evolution of the binding energy with the X energy for X$^-$ (red triangles), XX (black circles), and X$^+$ (blue diamonds). }
\end{figure}

%%%%%%%
%%%%%%%
%\vspace{1em}
%\noindent\textbf{\textit{Spectral characteristics of single dots.}} 
Figure~\ref{multiX}(a) shows the typical PL spectrum of a single dot. Four emission lines are observed, and assigned, from the high-energy side, as X, X$^+$, XX, and X$^-$, where X$^{+(-)}$ refers to positively (negatively) charged excitons. These assignments are based on the measurement of the excitation power dependence of the emission lines (Fig.~\ref{multiX}(b)), where almost linear and quadratic behaviors were observed for X and XX, respectively. The assignment of X$^{+}$ and X$^{-}$ is further supported by an optical orientation measurement, as described later.

Figure~\ref{multiX}(c) shows the binding energy of each exciton complex as a function of X energy. Here the binding energy is defined as the energy difference between X and the exciton complex. It reveals that, with increasing X energy, the binding energy of X$^-$ increases, that of X$^+$ decreases, and that of XX has intermediate values. The mirror symmetric evolution of X$^-$ and X$^+$ is induced by the mean-field contribution to exciton charging \cite{Abbarchi2010}. %, as was also observed in droplet epitaxial GaAs dots. 
Note that the spectral profile of exciton complexes is known to show a sensitive dependence on dot structure \cite{Mlinar_PRB09}. Therefore, the observation of a clear and less dispersive evolution in the binding energy suggests that the shape and other microscopic parameters of dots with a given size are almost identical. This is likely to be due to the kinetically limited formation of dots for droplet epitaxy. Consequently, dots on (111) substrates become rather symmetric as microscopic randomness is effectively suppressed.

\begin{figure}
\includegraphics[width=9cm]{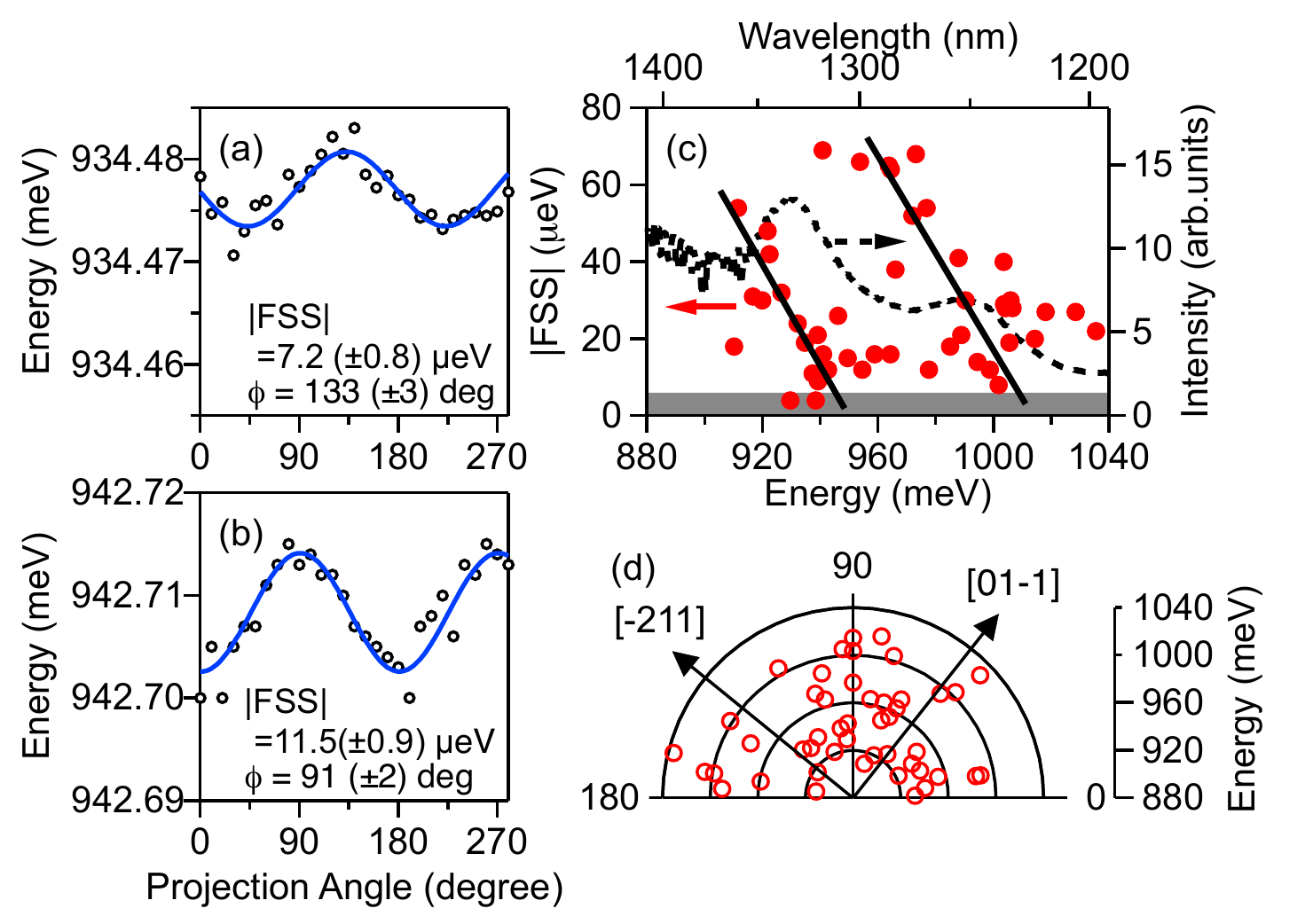}%
\caption{\label{fss}(color online) (a, b) Energy shifts of the X line with rotating the axis of linear polarization for two dots. The blue lines are sinusoidal fits to the data. (c) Absolute value of FSS measured on single QDs (red circles). The broken line shows the ensemble PL spectrum. The solid lines are guides to the eye. (d) Polarization axis with respect to the emission energy. Two orthogonal in-plane axes [01-1] and [-211] are shown by arrows. The equivalent directions appear with every 120$^{\circ}$ rotation.}
\end{figure}

%%%%%%%
%%%%%%%
%\vspace{1em}
%\noindent\textbf{\textit{Distribution of fine-structure splittings.}} 
Figures~\ref{fss}(a) and (b) show the evolution of the X peak energy when the linear polarization axis is rotated for two dots. The precise quantification of FSS is based on sinusoidal fitting to these evolutions, where the magnitude (absolute value) of FSS is defined as the amplitude of sine curves, and the polarization axis, $\phi$, is defined as the first maximum phase. Thus, the angle of $\phi$ corresponds to the polarization axis of the high-energy X line among two split lines with orthogonal polarization. As shown by Figs.~\ref{fss}(a) and (b), both the FSS magnitude and polarization axis differ dot by dot. 

The statistical results for FSS over 50 dots are summarized in Fig.~\ref{fss}(c), where the FSS magnitude is plotted as a function of X energy. The FSS ranges between 70 and 3~$\mu$eV, where the minimum value is smaller than the error width of the present analysis (4~$\mu$eV, shown by the shaded region). The FSS average value is 25~$\mu$eV, which is considerably smaller than those of SK grown QDs in the telecom band \cite{Cade_PRB06,Chauvin2006,Sapienza2013}. Note that two families of QDs with different monolayer heights are present in Fig.~\ref{fss}(c), as shown by the two peaks in the ensemble spectrum (broken line). Each family with a given height reveals a trend where the FSS decreases as the energy increases. This implies that high-energy QDs have a smaller in-plane size and higher lateral symmetry. %Similar findings
The impact of the lateral-size reduction on FSS minimization was also confirmed in previous study on FSS control by high-temperature annealing \cite{Lagbein_PRB04,*Young_PRB05}. 

%The statistical results of FSS over 50 dots are summarized in Fig.~\ref{fss}(c), where the magnitude of FSS is plotted as a function of X energy. It spreads between 70~$\mu$eV and 3~$\mu$eV, where the minimum value is smaller than the error width of the present analysis (4~$\mu$eV, shown by the shaded region). The average value of FSS is found to be 25~$\mu$eV, which is considerably smaller than those of SK grown QDs in telecom band \cite{Cade_PRB06,Chauvin2006,Sapienza2013}. Note that two families of QDs with different monolayer heights are present in Fig.~\ref{fss}(c), as shown by two peaks in the ensemble spectrum. Each family with a given height reveals a trend of decreasing FSS with increasing energy. This implies that high-energy QDs have smaller in-plane size and higher lateral symmetry. Similar findings were confirmed in previous works on the FSS control by high-temperature annealing \cite{Lagbein_PRB04,*Young_PRB05}. 

Figure~\ref{fss}(d) shows the direction of the polarization axis with respect to the X energy. They are randomly distributed, without showing significant correlations with the in-plane crystallographic axes. The absence of a preferential direction in the (111) plane suggests a high probability of finding QDs with negligible FSS over a broad spectral range. 

%Figure~\ref{fss}(d) shows the direction of polarization axis with respect to X energy. They are randomly distributed, without showing significant correlations with the in-plane crystallographic axes. The absence of preferential direction in the (111) plane suggests a high probability of finding QDs with negligible FSS in a broad spectral range. 

\begin{figure}
\includegraphics[width=8cm]{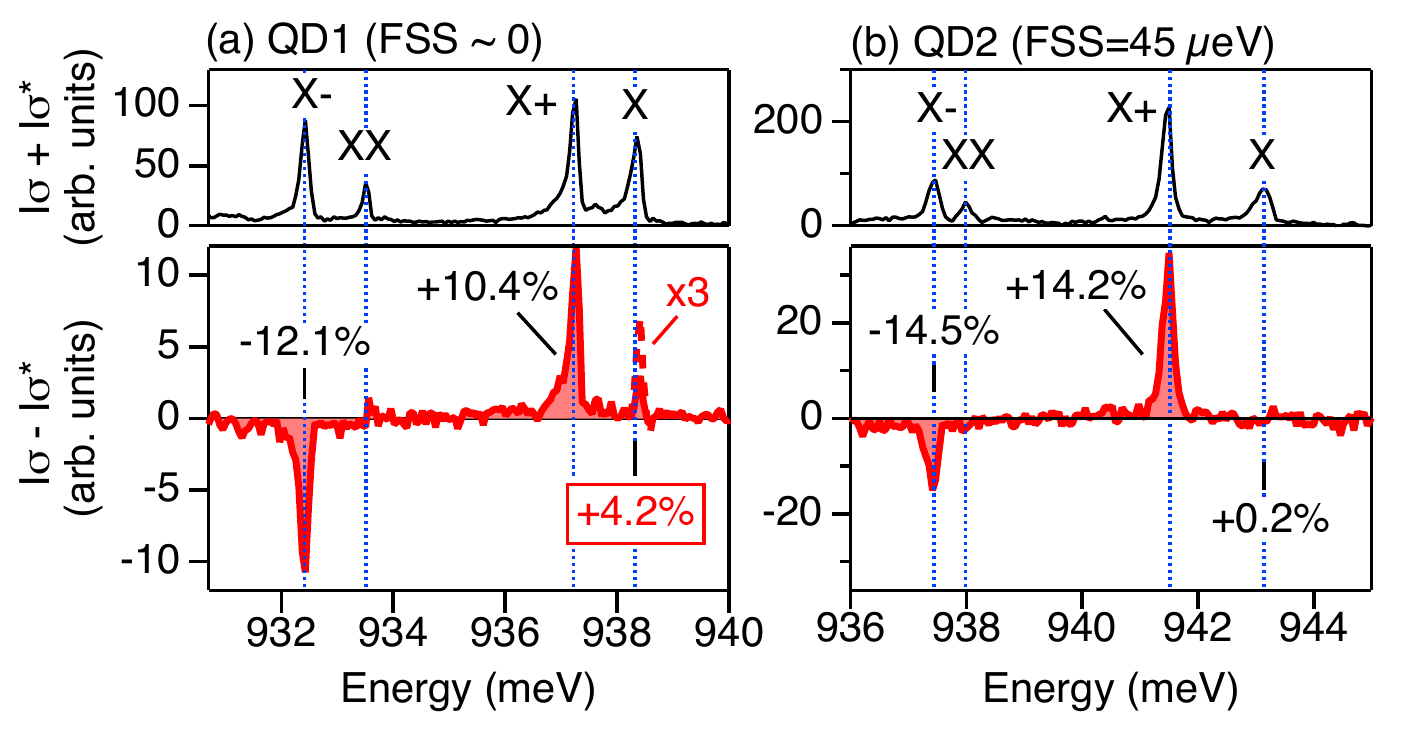}%
\caption{\label{dop}(color online) Circularly-polarized emission signals with non-resonant circularly-polarized pump at a wavelength of 705~nm for (a) QD without a detectable FSS, and (b) QD with a finite value of FSS. Upper panels show unpolarized spectra. Lower panels show differential spectra between co-circular signals and cross-circular signals. The degree of circular polarization, $(I_{\sigma} - I_{\sigma ^*})/(I_{\sigma} + I_{\sigma ^*})$, of each emission line is also shown.}
 \end{figure}

%%%%%%%
%%%%%%%
%\vspace{1em}
%\noindent\textbf{\textit{Optical orientation -- a proof of zero splitting.}} 
The presence of dots with effectively zero FSS can be confirmed by measuring circularly polarized emission signals. In the presence of a finite FSS value, the polarization state of the emission light is expected to oscillate temporally between the left- and right-handed circular polarizations, where the oscillation period is determined by the inverse of the FSS. Therefore, in time-integrated experiments we cannot observe a high degree of circular polarization. By contrast, in the absence of FSS, circular polarization remains in time-integrated signals after circular injection. Thus, the observation of circular polarization provides a sufficient condition for exciton degeneracy. The measurement principle is analogous to the well-known Hanle measurement, and was used to monitor FSS cancellation by an electric field \cite{Kowalik_APL07}. To avoid the effect of dynamic nuclear polarization \cite{Belhadj_PRL09}, we set the excitation power at a sufficiently low level, where the average exciton population in the dot was $\sim0.5$.

%The presence of dots with effectively zero FSS can be confirmed by the measurement of circularly polarized emission signals. In the presence of a finite value of FSS, the polarization state of emission light is expected to temporally oscillate between the left- and right-handed circular polarizations, where the oscillation interval is determined by the inverse of FSS. For time-integrated experiments, one therefore cannot observe a high degree of circular polarization. In the absence of FSS, in contrast, circular polarization remains in time-integrated signals after circular injection. Thus, the observation of circular polarization provides a sufficient condition of exciton degeneracy. The measurement principle is analogous to the well-known Hanle measurement, and was used to monitor FSS cancellation by an electric field \cite{Kowalik_APL07}. To avoid the effect of dynamic nuclear polarization \cite{Belhadj_PRL09}, we set the excitation power to a sufficiently low level. 

Figures~\ref{dop}(a) and (b) show optical orientation results for a selected dot without a detectable FSS ($<4\,\mu$eV, QD1) and for a dot with a significant FSS ($\sim 45 \, \mu$eV, QD2), respectively. The upper panels show unpolarized spectra, $I_{\sigma} + I_{\sigma ^*}$, where $I_{\sigma} \, (I_{\sigma ^*})$ is the emission intensity with co-circular (cross-circular) polarization with respect to the excitation light. The lower panels show the differential spectra, $I_{\sigma} - I_{\sigma ^*}$, where a pronounced positive peak appears for the X line of QD1, but disappears for that of QD2. The positive degree of polarization for the X line of QD1 ($+4.2$\%) is evidence of the degenerate exciton states. By contrast, the XX line does not exhibit a significant polarization in both QD1 and QD2, because the transition from XX comprises two routes with orthogonal polarizations. The other spectral lines follow well-known dynamics: X$^+$ shows a positive degree of polarization, which is due to spin-polarized electron injection. X$^-$ shows a negative degree of polarization, which is accompanied by the spin-flip relaxation of electrons \cite{Cortez_PRL02,Bracker_PRL05}.

%The results of optical orientation are shown for a selected dot which does not have a detectable FSS ($<4\,\mu$eV, QD1) in Fig.~\ref{dop}(a), and for a dot with a significant FSS ($\sim 29 \, \mu$eV, QD2) in Fig.~\ref{dop}(b). The upper panels show un-polarized spectra, $I_{\sigma} + I_{\sigma ^*}$, where $I_{\sigma} \, (I_{\sigma ^*})$ is the emission intensity with co-circular (cross-circular) polarization with respect to the excitation light. The lower panels show the differential spectra, $I_{\sigma} - I_{\sigma ^*}$, where a pronounced positive peak appears for the X line of QD1, but it disappears for that of QD2. The positive degree of polarization ($+4$ \% for QD1) is an evidence of the degenerate exciton states. 

%The signals for the other spectral lines follow well-known dynamics: X$^+$ shows a positive degree of polarization, which is due to spin-polarized electron injection. X$^-$ shows a negative degree of polarization, which is accompanied by the spin-flip relaxation of electrons \cite{Cortez_PRL02,Bracker_PRL05}. In our sample, the X$^-$ polarization strongly depends on the dot ($-11$~\% for QD1, and $-7$~\% for QD2). 
%XX does not exhibit a significant polarization, since the transition from XX comprises two routes with orthogonal polarizations. 
Note that only a few dots exhibit circular polarization for the X line. A rough estimation of the probability of finding dots with a circularly polarized X line is $\sim 2$\%, which agrees with the ratio of the natural width of our dots (0.5~$\mu$eV) divided by the distribution of FSS (25~$\mu$eV). This small probability reflects the relatively long emission lifetime of telecom-wavelngth dots as compared with that of visible-wavelength dots. Optical orientation therefore serves as an efficient way to select dots suited for entangled photon generation. Although it is not essentially difficult to find dots with effectively zero FSS, a combination of external tuning protocols is also beneficial, where only a small tuning range is required to reach the optimum conditions in our dots. 

%The signals for the other spectral lines follow well-known dynamics: X$^+$ shows a positive degree of polarization, which is due to spin-polarized electron injection. X$^-$ shows a negative degree of polarization, which is accompanied with spin-flip relaxation of electrons \cite{Cortez_PRL02,Bracker_PRL05}. In our sample, the X$^-$ polarization strongly depends on dot ($-11$~\% for QD1, and $-7$~\% for QD2). XX does not show a significant polarization, since transition from XX comprises two routes with orthogonal polarizations. Note that the majority of dots does not reveal a circular polarization degree for the X line, and a few numbers of dots were confirmed to reveal it. A rough estimation of the probability of finding dots with the circularly-polarized X line is $\sim 2$~\%, which is in agreement with the ratio of the natural width of our dots (0.5~$\mu$eV) divided by the distribution of FSS (25~$\mu$eV). Optical orientation therefore serves as an efficient method to select dots suited for entangled photon generation. Though it is not essentially difficult to find dots with effectively-zero FSS, the co-usage of external tuning protocols is also beneficial, where a small tuning range is only required to reach optimum conditions in our dots. 

%%%%%%%
%%%%%%%
%\vspace{1em}
%\noindent\textbf{\textit{Discussion -- Origin of symmetry conservation for dots on (111) surfaces.}} 
Recent theoretical attempts based on first-principle calculations suggest the influence of atomic-scale symmetry breaking on the emergence of FSS \cite{Bester_PRB03,*He_PRL08}. Such microscopic asymmetry comes from interfacial randomness at hetero surfaces and compositional fluctuations inside and outside dots. 
%\textcolor{green}{In addition, FSS is further enhanced by an anisotropic piezoelectric field caused by the local strain distribution. }
It is noteworthy that the distribution of the measured FSS in our dots is smaller than that theoretically predicted for telecom dots on (100) surfaces, in which a perfectly symmetric shape was assumed \cite{Goldmann_APL13}. This implies that the [111] grown dots are more stable against microscopic disorder than the [100] grown dots. 

We attribute the further reduction of FSS in the [111] grown dots compared with that of conventional [100] grown dots to two mechanisms. First, owing to the high surface stability of the (111) plane, the dots have atomically flat surfaces, which were demonstrated by transmission electron microscopy analysis \cite{Ha_APL14}. The smooth and abrupt interface also leads to the observation of distinct spectral multiplets in ensemble spectra (Fig.~\ref{ensemble}(a)). Thus, we expect the effect of interfacial randomness on FSS to be greatly suppressed compared with SK-grown (100) dots. %
%
%Recent theoretical attempts based on first-principle calculations suggest the influence of atomic-scale symmetry braking on the emergence of FSS \cite{Bester_PRB03,*He_PRL08}. Such microscopic asymmetry comes with interfacial randomness at hetero surfaces and composition fluctuations inside and outside dots. Remarkably, the distribution of measured FSS in our dots is smaller than that theoretically predicted for telecom dots on (100) surfaces, in which a perfectly symmetric shape was assumed \cite{Goldmann_APL13}. We attribute two mechanism to further reduction of FSS in the [111] grown dots, compared with conventional [100] grown dots. First, owing to the high stability of the (111) plane, the dots have atomically-flat top and bottom surfaces. The presence of abrupt interfaces was demonstrated by transmission electron microscopy analysis \cite{Ha_APL14}. The flat surface also leads to the observation of distinct spectral multiplets in ensemble spectra (Fig.~\ref{ensemble}(a)). Thus, we expect that the effect of interfacial asymmetry is greatly suppressed, compared with SK-grown (100) dots. 
%
Second, in zinc-blend compound semiconductors, the piezoelectric field direction is along the [111] polar axis, which coincides with the vertical growth direction in our system. Thus, a strain field does not induce any great reduction in lateral symmetry. The (111) surface is thus an ideal substrate for the growth of symmetric dots, where both geometrical (shape) symmetry and atomic-scale symmetry are well conserved. 

%Second, the direction of a piezo-electric field is along the [111] polar axis, which coincides with a vertical direction in our system. Thus, a strain field does not majorly induce lateral symmetry reduction. The (111) surface is thus an ideal substrate for the growth of symmetric dots, where both geometrical (shape) symmetry and atomic-scale symmetry are well conserved. 

%\vspace{1em}
%\noindent\textbf{\textit{Conclusions.}} 
In summary, we have presented measurements of minimized FSS in telecom-wavelength InAs QDs on an InP(111)A substrate prepared by droplet epitaxy. Polarization-resolved PL measurements were performed to examine the FSS distribution. Resolution-limited splittings (smaller than 4 $\mu$eV) were confirmed. The random distribution of the polarization axis made it possible to find symmetric dots over a wide spectral range. Thus our InAs/InAlAs dots on (111) substrates can play a crucial role in quantum information processing as an efficient entangled photon source that can work in telecom fiber networks.

\newpage
\bibliography{telecom}

\providecommand{\noopsort}[1]{}\providecommand{\singleletter}[1]{#1}%
\begin{thebibliography}{60}%
\makeatletter
\providecommand \@ifxundefined [1]{%
 \@ifx{#1\undefined}
}%
\providecommand \@ifnum [1]{%
 \ifnum #1\expandafter \@firstoftwo
 \else \expandafter \@secondoftwo
 \fi
}%
\providecommand \@ifx [1]{%
 \ifx #1\expandafter \@firstoftwo
 \else \expandafter \@secondoftwo
 \fi
}%
\providecommand \natexlab [1]{#1}%
\providecommand \enquote  [1]{``#1''}%
\providecommand \bibnamefont  [1]{#1}%
\providecommand \bibfnamefont [1]{#1}%
\providecommand \citenamefont [1]{#1}%
\providecommand \href@noop [0]{\@secondoftwo}%
\providecommand \href [0]{\begingroup \@sanitize@url \@href}%
\providecommand \@href[1]{\@@startlink{#1}\@@href}%
\providecommand \@@href[1]{\endgroup#1\@@endlink}%
\providecommand \@sanitize@url [0]{\catcode `\\12\catcode `\$12\catcode
  `\&12\catcode `\#12\catcode `\^12\catcode `\_12\catcode `\%12\relax}%
\providecommand \@@startlink[1]{}%
\providecommand \@@endlink[0]{}%
\providecommand \url  [0]{\begingroup\@sanitize@url \@url }%
\providecommand \@url [1]{\endgroup\@href {#1}{\urlprefix }}%
\providecommand \urlprefix  [0]{URL }%
\providecommand \Eprint [0]{\href }%
\providecommand \doibase [0]{http://dx.doi.org/}%
\providecommand \selectlanguage [0]{\@gobble}%
\providecommand \bibinfo  [0]{\@secondoftwo}%
\providecommand \bibfield  [0]{\@secondoftwo}%
\providecommand \translation [1]{[#1]}%
\providecommand \BibitemOpen [0]{}%
\providecommand \bibitemStop [0]{}%
\providecommand \bibitemNoStop [0]{.\EOS\space}%
\providecommand \EOS [0]{\spacefactor3000\relax}%
\providecommand \BibitemShut  [1]{\csname bibitem#1\endcsname}%
\let\auto@bib@innerbib\@empty
%</preamble>
\bibitem [{\citenamefont {Zeilinger}(1999)}]{Zeillinger_RMP99}%
  \BibitemOpen
  \bibfield  {author} {\bibinfo {author} {\bibfnamefont {A.}~\bibnamefont
  {Zeilinger}},\ }\bibfield  {title} {\enquote {\bibinfo {title} {Experiment
  and the foundations of quantum physics},}\ }\href {\doibase
  10.1103/RevModPhys.71.S288} {\bibfield  {journal} {\bibinfo  {journal} {Rev.
  Mod. Phys.}\ }\textbf {\bibinfo {volume} {71}},\ \bibinfo {pages}
  {S288--S297} (\bibinfo {year} {1999})}\BibitemShut {NoStop}%
\bibitem [{\citenamefont {Gisin}\ \emph {et~al.}(2002)\citenamefont {Gisin},
  \citenamefont {Ribordy}, \citenamefont {Tittel},\ and\ \citenamefont
  {Zbinden}}]{Gisin_RMP02}%
  \BibitemOpen
  \bibfield  {author} {\bibinfo {author} {\bibfnamefont {N.}~\bibnamefont
  {Gisin}}, \bibinfo {author} {\bibfnamefont {G.}~\bibnamefont {Ribordy}},
  \bibinfo {author} {\bibfnamefont {W.}~\bibnamefont {Tittel}}, \ and\ \bibinfo
  {author} {\bibfnamefont {H.}~\bibnamefont {Zbinden}},\ }\bibfield  {title}
  {\enquote {\bibinfo {title} {Quantum cryptography},}\ }\href {\doibase
  10.1103/RevModPhys.74.145} {\bibfield  {journal} {\bibinfo  {journal} {Rev.
  Mod. Phys.}\ }\textbf {\bibinfo {volume} {74}},\ \bibinfo {pages} {145--195}
  (\bibinfo {year} {2002})}\BibitemShut {NoStop}%
\bibitem [{\citenamefont {Benson}\ \emph {et~al.}(2000)\citenamefont {Benson},
  \citenamefont {Santori}, \citenamefont {Pelton},\ and\ \citenamefont
  {Yamamoto}}]{Benson_PRL00}%
  \BibitemOpen
  \bibfield  {author} {\bibinfo {author} {\bibfnamefont {O.}~\bibnamefont
  {Benson}}, \bibinfo {author} {\bibfnamefont {C.}~\bibnamefont {Santori}},
  \bibinfo {author} {\bibfnamefont {M.}~\bibnamefont {Pelton}}, \ and\ \bibinfo
  {author} {\bibfnamefont {Y.}~\bibnamefont {Yamamoto}},\ }\bibfield  {title}
  {\enquote {\bibinfo {title} {Regulated and entangled photons from a single
  quantum dot},}\ }\href {\doibase 10.1103/PhysRevLett.84.2513} {\bibfield
  {journal} {\bibinfo  {journal} {Phys. Rev. Lett.}\ }\textbf {\bibinfo
  {volume} {84}},\ \bibinfo {pages} {2513--2516} (\bibinfo {year}
  {2000})}\BibitemShut {NoStop}%
\bibitem [{\citenamefont {Takemoto}\ \emph {et~al.}(2010)\citenamefont
  {Takemoto}, \citenamefont {Nambu}, \citenamefont {Miyazawa}, \citenamefont
  {Wakui}, \citenamefont {Hirose}, \citenamefont {Usuki}, \citenamefont
  {Takatsu}, \citenamefont {Yokoyama}, \citenamefont {Yoshino}, \citenamefont
  {Tomita}, \citenamefont {Yorozu}, \citenamefont {Sakuma},\ and\ \citenamefont
  {Arakawa}}]{Takemoto_APEX10}%
  \BibitemOpen
  \bibfield  {author} {\bibinfo {author} {\bibfnamefont {K.}~\bibnamefont
  {Takemoto}}, \bibinfo {author} {\bibfnamefont {Y.}~\bibnamefont {Nambu}},
  \bibinfo {author} {\bibfnamefont {T.}~\bibnamefont {Miyazawa}}, \bibinfo
  {author} {\bibfnamefont {K.}~\bibnamefont {Wakui}}, \bibinfo {author}
  {\bibfnamefont {S.}~\bibnamefont {Hirose}}, \bibinfo {author} {\bibfnamefont
  {T.}~\bibnamefont {Usuki}}, \bibinfo {author} {\bibfnamefont
  {M.}~\bibnamefont {Takatsu}}, \bibinfo {author} {\bibfnamefont
  {N.}~\bibnamefont {Yokoyama}}, \bibinfo {author} {\bibfnamefont
  {K.}~\bibnamefont {Yoshino}}, \bibinfo {author} {\bibfnamefont
  {A.}~\bibnamefont {Tomita}}, \bibinfo {author} {\bibfnamefont
  {S.}~\bibnamefont {Yorozu}}, \bibinfo {author} {\bibfnamefont
  {Y.}~\bibnamefont {Sakuma}}, \ and\ \bibinfo {author} {\bibfnamefont
  {Y.}~\bibnamefont {Arakawa}},\ }\bibfield  {title} {\enquote {\bibinfo
  {title} {Transmission experiment of quantum keys over 50 km using
  high-performance quantum-dot single-photon source at 1.5 $\mu$m
  wavelength},}\ }\href {http://stacks.iop.org/1882-0786/3/i=9/a=092802}
  {\bibfield  {journal} {\bibinfo  {journal} {Appl. Phys. Express}\ }\textbf
  {\bibinfo {volume} {3}},\ \bibinfo {pages} {092802} (\bibinfo {year}
  {2010})}\BibitemShut {NoStop}%
\bibitem [{\citenamefont {Bennett}\ \emph {et~al.}(1993)\citenamefont
  {Bennett}, \citenamefont {Brassard}, \citenamefont {Cr\'epeau}, \citenamefont
  {Jozsa}, \citenamefont {Peres},\ and\ \citenamefont
  {Wootters}}]{Bennett_PRL93}%
  \BibitemOpen
  \bibfield  {author} {\bibinfo {author} {\bibfnamefont {C.~H.}\ \bibnamefont
  {Bennett}}, \bibinfo {author} {\bibfnamefont {G.}~\bibnamefont {Brassard}},
  \bibinfo {author} {\bibfnamefont {C.}~\bibnamefont {Cr\'epeau}}, \bibinfo
  {author} {\bibfnamefont {R.}~\bibnamefont {Jozsa}}, \bibinfo {author}
  {\bibfnamefont {A.}~\bibnamefont {Peres}}, \ and\ \bibinfo {author}
  {\bibfnamefont {W.~K.}\ \bibnamefont {Wootters}},\ }\bibfield  {title}
  {\enquote {\bibinfo {title} {Teleporting an unknown quantum state via dual
  classical and {E}instein-{P}odolsky-{R}osen channels},}\ }\href {\doibase
  10.1103/PhysRevLett.70.1895} {\bibfield  {journal} {\bibinfo  {journal}
  {Phys. Rev. Lett.}\ }\textbf {\bibinfo {volume} {70}},\ \bibinfo {pages}
  {1895--1899} (\bibinfo {year} {1993})}\BibitemShut {NoStop}%
\bibitem [{\citenamefont {\ifmmode~\dot{Z}\else \.{Z}\fi{}ukowski}\ \emph
  {et~al.}(1993)\citenamefont {\ifmmode~\dot{Z}\else \.{Z}\fi{}ukowski},
  \citenamefont {Zeilinger}, \citenamefont {Horne},\ and\ \citenamefont
  {Ekert}}]{Zukowski_PRL93}%
  \BibitemOpen
  \bibfield  {author} {\bibinfo {author} {\bibfnamefont {M.}~\bibnamefont
  {\ifmmode~\dot{Z}\else \.{Z}\fi{}ukowski}}, \bibinfo {author} {\bibfnamefont
  {A.}~\bibnamefont {Zeilinger}}, \bibinfo {author} {\bibfnamefont {M.~A.}\
  \bibnamefont {Horne}}, \ and\ \bibinfo {author} {\bibfnamefont {A.~K.}\
  \bibnamefont {Ekert}},\ }\bibfield  {title} {\enquote {\bibinfo {title}
  {{``Event-ready-detectors''} {B}ell experiment via entanglement swapping},}\
  }\href {\doibase 10.1103/PhysRevLett.71.4287} {\bibfield  {journal} {\bibinfo
   {journal} {Phys. Rev. Lett.}\ }\textbf {\bibinfo {volume} {71}},\ \bibinfo
  {pages} {4287--4290} (\bibinfo {year} {1993})}\BibitemShut {NoStop}%
\bibitem [{\citenamefont {Bose}, \citenamefont {Vedral},\ and\ \citenamefont
  {Knight}(1998)}]{Bose_PRA98}%
  \BibitemOpen
  \bibfield  {author} {\bibinfo {author} {\bibfnamefont {S.}~\bibnamefont
  {Bose}}, \bibinfo {author} {\bibfnamefont {V.}~\bibnamefont {Vedral}}, \ and\
  \bibinfo {author} {\bibfnamefont {P.~L.}\ \bibnamefont {Knight}},\ }\bibfield
   {title} {\enquote {\bibinfo {title} {Multiparticle generalization of
  entanglement swapping},}\ }\href {\doibase 10.1103/PhysRevA.57.822}
  {\bibfield  {journal} {\bibinfo  {journal} {Phys. Rev. A}\ }\textbf {\bibinfo
  {volume} {57}},\ \bibinfo {pages} {822--829} (\bibinfo {year}
  {1998})}\BibitemShut {NoStop}%
\bibitem [{\citenamefont {Pan}\ \emph {et~al.}(1998)\citenamefont {Pan},
  \citenamefont {Bouwmeester}, \citenamefont {Weinfurter},\ and\ \citenamefont
  {Zeilinger}}]{Pan_PRL98}%
  \BibitemOpen
  \bibfield  {author} {\bibinfo {author} {\bibfnamefont {J.-W.}\ \bibnamefont
  {Pan}}, \bibinfo {author} {\bibfnamefont {D.}~\bibnamefont {Bouwmeester}},
  \bibinfo {author} {\bibfnamefont {H.}~\bibnamefont {Weinfurter}}, \ and\
  \bibinfo {author} {\bibfnamefont {A.}~\bibnamefont {Zeilinger}},\ }\bibfield
  {title} {\enquote {\bibinfo {title} {Experimental entanglement swapping:
  Entangling photons that never interacted},}\ }\href {\doibase
  10.1103/PhysRevLett.80.3891} {\bibfield  {journal} {\bibinfo  {journal}
  {Phys. Rev. Lett.}\ }\textbf {\bibinfo {volume} {80}},\ \bibinfo {pages}
  {3891--3894} (\bibinfo {year} {1998})}\BibitemShut {NoStop}%
\bibitem [{\citenamefont {Gammon}\ \emph {et~al.}(1996)\citenamefont {Gammon},
  \citenamefont {Snow}, \citenamefont {Shanabrook}, \citenamefont {Katzer},\
  and\ \citenamefont {Park}}]{Gammon_PRL96}%
  \BibitemOpen
  \bibfield  {author} {\bibinfo {author} {\bibfnamefont {D.}~\bibnamefont
  {Gammon}}, \bibinfo {author} {\bibfnamefont {E.~S.}\ \bibnamefont {Snow}},
  \bibinfo {author} {\bibfnamefont {B.~V.}\ \bibnamefont {Shanabrook}},
  \bibinfo {author} {\bibfnamefont {D.~S.}\ \bibnamefont {Katzer}}, \ and\
  \bibinfo {author} {\bibfnamefont {D.}~\bibnamefont {Park}},\ }\bibfield
  {title} {\enquote {\bibinfo {title} {Fine structure splitting in the optical
  spectra of single {GaAs} quantum dots},}\ }\href {\doibase
  10.1103/PhysRevLett.76.3005} {\bibfield  {journal} {\bibinfo  {journal}
  {Phys. Rev. Lett.}\ }\textbf {\bibinfo {volume} {76}},\ \bibinfo {pages}
  {3005--3008} (\bibinfo {year} {1996})}\BibitemShut {NoStop}%
\bibitem [{\citenamefont {Bayer}\ \emph {et~al.}(1999)\citenamefont {Bayer},
  \citenamefont {Kuther}, \citenamefont {Forchel}, \citenamefont {Gorbunov},
  \citenamefont {Timofeev}, \citenamefont {Sch\"afer}, \citenamefont
  {Reithmaier}, \citenamefont {Reinecke},\ and\ \citenamefont
  {Walck}}]{Bayer_PRL99}%
  \BibitemOpen
  \bibfield  {author} {\bibinfo {author} {\bibfnamefont {M.}~\bibnamefont
  {Bayer}}, \bibinfo {author} {\bibfnamefont {A.}~\bibnamefont {Kuther}},
  \bibinfo {author} {\bibfnamefont {A.}~\bibnamefont {Forchel}}, \bibinfo
  {author} {\bibfnamefont {A.}~\bibnamefont {Gorbunov}}, \bibinfo {author}
  {\bibfnamefont {V.~B.}\ \bibnamefont {Timofeev}}, \bibinfo {author}
  {\bibfnamefont {F.}~\bibnamefont {Sch\"afer}}, \bibinfo {author}
  {\bibfnamefont {J.~P.}\ \bibnamefont {Reithmaier}}, \bibinfo {author}
  {\bibfnamefont {T.~L.}\ \bibnamefont {Reinecke}}, \ and\ \bibinfo {author}
  {\bibfnamefont {S.~N.}\ \bibnamefont {Walck}},\ }\bibfield  {title} {\enquote
  {\bibinfo {title} {Electron and hole $\mathit{g}$ factors and exchange
  interaction from studies of the exciton fine structure in
  {In}$_{0.60}${Ga}$_{0.40}${As} quantum dots},}\ }\href {\doibase
  10.1103/PhysRevLett.82.1748} {\bibfield  {journal} {\bibinfo  {journal}
  {Phys. Rev. Lett.}\ }\textbf {\bibinfo {volume} {82}},\ \bibinfo {pages}
  {1748--1751} (\bibinfo {year} {1999})}\BibitemShut {NoStop}%
\bibitem [{\citenamefont {Kulakovskii}\ \emph {et~al.}(1999)\citenamefont
  {Kulakovskii}, \citenamefont {Bacher}, \citenamefont {Weigand}, \citenamefont
  {K\"ummell}, \citenamefont {Forchel}, \citenamefont {Borovitskaya},
  \citenamefont {Leonardi},\ and\ \citenamefont {Hommel}}]{Kulakovskii_PRL99}%
  \BibitemOpen
  \bibfield  {author} {\bibinfo {author} {\bibfnamefont {V.~D.}\ \bibnamefont
  {Kulakovskii}}, \bibinfo {author} {\bibfnamefont {G.}~\bibnamefont {Bacher}},
  \bibinfo {author} {\bibfnamefont {R.}~\bibnamefont {Weigand}}, \bibinfo
  {author} {\bibfnamefont {T.}~\bibnamefont {K\"ummell}}, \bibinfo {author}
  {\bibfnamefont {A.}~\bibnamefont {Forchel}}, \bibinfo {author} {\bibfnamefont
  {E.}~\bibnamefont {Borovitskaya}}, \bibinfo {author} {\bibfnamefont
  {K.}~\bibnamefont {Leonardi}}, \ and\ \bibinfo {author} {\bibfnamefont
  {D.}~\bibnamefont {Hommel}},\ }\bibfield  {title} {\enquote {\bibinfo {title}
  {Fine structure of biexciton emission in symmetric and asymmetric {CdSe/ZnSe}
  single quantum dots},}\ }\href {\doibase 10.1103/PhysRevLett.82.1780}
  {\bibfield  {journal} {\bibinfo  {journal} {Phys. Rev. Lett.}\ }\textbf
  {\bibinfo {volume} {82}},\ \bibinfo {pages} {1780--1783} (\bibinfo {year}
  {1999})}\BibitemShut {NoStop}%
\bibitem [{\citenamefont {Seguin}\ \emph {et~al.}(2005)\citenamefont {Seguin},
  \citenamefont {Schliwa}, \citenamefont {Rodt}, \citenamefont {P{\"o}tschke},
  \citenamefont {Pohl},\ and\ \citenamefont {Bimberg}}]{Seguin2005}%
  \BibitemOpen
  \bibfield  {author} {\bibinfo {author} {\bibfnamefont {R.}~\bibnamefont
  {Seguin}}, \bibinfo {author} {\bibfnamefont {A.}~\bibnamefont {Schliwa}},
  \bibinfo {author} {\bibfnamefont {S.}~\bibnamefont {Rodt}}, \bibinfo {author}
  {\bibfnamefont {K.}~\bibnamefont {P{\"o}tschke}}, \bibinfo {author}
  {\bibfnamefont {U.~W.}\ \bibnamefont {Pohl}}, \ and\ \bibinfo {author}
  {\bibfnamefont {D.}~\bibnamefont {Bimberg}},\ }\bibfield  {title} {\enquote
  {\bibinfo {title} {Size-dependent fine-structure splitting in self-organized
  {InAs}/{GaAs} quantum dots},}\ }\href {\doibase
  10.1103/PhysRevLett.95.257402} {\bibfield  {journal} {\bibinfo  {journal}
  {Phys. Phys. Lett.}\ }\textbf {\bibinfo {volume} {95}},\ \bibinfo {pages}
  {257402} (\bibinfo {year} {2005})}\BibitemShut {NoStop}%
\bibitem [{\citenamefont {Abbarchi}\ \emph {et~al.}(2008)\citenamefont
  {Abbarchi}, \citenamefont {Mastrandrea}, \citenamefont {Kuroda},
  \citenamefont {Mano}, \citenamefont {Sakoda}, \citenamefont {Koguchi},
  \citenamefont {Sanguinetti}, \citenamefont {Vinattieri},\ and\ \citenamefont
  {Gurioli}}]{Marco_PRB08}%
  \BibitemOpen
  \bibfield  {author} {\bibinfo {author} {\bibfnamefont {M.}~\bibnamefont
  {Abbarchi}}, \bibinfo {author} {\bibfnamefont {C.~A.}\ \bibnamefont
  {Mastrandrea}}, \bibinfo {author} {\bibfnamefont {T.}~\bibnamefont {Kuroda}},
  \bibinfo {author} {\bibfnamefont {T.}~\bibnamefont {Mano}}, \bibinfo {author}
  {\bibfnamefont {K.}~\bibnamefont {Sakoda}}, \bibinfo {author} {\bibfnamefont
  {N.}~\bibnamefont {Koguchi}}, \bibinfo {author} {\bibfnamefont
  {S.}~\bibnamefont {Sanguinetti}}, \bibinfo {author} {\bibfnamefont
  {A.}~\bibnamefont {Vinattieri}}, \ and\ \bibinfo {author} {\bibfnamefont
  {M.}~\bibnamefont {Gurioli}},\ }\bibfield  {title} {\enquote {\bibinfo
  {title} {Exciton fine structure in strain-free
  {GaAs/Al}$_{0.3}${Ga}$_{0.7}${As} quantum dots: Extrinsic effects},}\ }\href
  {\doibase 10.1103/PhysRevB.78.125321} {\bibfield  {journal} {\bibinfo
  {journal} {Phys. Rev. B}\ }\textbf {\bibinfo {volume} {78}},\ \bibinfo
  {pages} {125321} (\bibinfo {year} {2008})}\BibitemShut {NoStop}%
\bibitem [{\citenamefont {Santori}\ \emph {et~al.}(2002)\citenamefont
  {Santori}, \citenamefont {Fattal}, \citenamefont {Pelton}, \citenamefont
  {Solomon},\ and\ \citenamefont {Yamamoto}}]{Santori_PRB02}%
  \BibitemOpen
  \bibfield  {author} {\bibinfo {author} {\bibfnamefont {C.}~\bibnamefont
  {Santori}}, \bibinfo {author} {\bibfnamefont {D.}~\bibnamefont {Fattal}},
  \bibinfo {author} {\bibfnamefont {M.}~\bibnamefont {Pelton}}, \bibinfo
  {author} {\bibfnamefont {G.~S.}\ \bibnamefont {Solomon}}, \ and\ \bibinfo
  {author} {\bibfnamefont {Y.}~\bibnamefont {Yamamoto}},\ }\bibfield  {title}
  {\enquote {\bibinfo {title} {Polarization-correlated photon pairs from a
  single quantum dot},}\ }\href {\doibase 10.1103/PhysRevB.66.045308}
  {\bibfield  {journal} {\bibinfo  {journal} {Phys. Rev. B}\ }\textbf {\bibinfo
  {volume} {66}},\ \bibinfo {pages} {045308} (\bibinfo {year}
  {2002})}\BibitemShut {NoStop}%
\bibitem [{\citenamefont {Langbein}\ \emph {et~al.}(2004)\citenamefont
  {Langbein}, \citenamefont {Borri}, \citenamefont {Woggon}, \citenamefont
  {Stavarache}, \citenamefont {Reuter},\ and\ \citenamefont
  {Wieck}}]{Lagbein_PRB04}%
  \BibitemOpen
  \bibfield  {author} {\bibinfo {author} {\bibfnamefont {W.}~\bibnamefont
  {Langbein}}, \bibinfo {author} {\bibfnamefont {P.}~\bibnamefont {Borri}},
  \bibinfo {author} {\bibfnamefont {U.}~\bibnamefont {Woggon}}, \bibinfo
  {author} {\bibfnamefont {V.}~\bibnamefont {Stavarache}}, \bibinfo {author}
  {\bibfnamefont {D.}~\bibnamefont {Reuter}}, \ and\ \bibinfo {author}
  {\bibfnamefont {A.~D.}\ \bibnamefont {Wieck}},\ }\bibfield  {title} {\enquote
  {\bibinfo {title} {Control of fine-structure splitting and biexciton binding
  in {In$_x$Ga$_{1-x}$As} quantum dots by annealing},}\ }\href {\doibase
  10.1103/PhysRevB.69.161301} {\bibfield  {journal} {\bibinfo  {journal} {Phys.
  Rev. B}\ }\textbf {\bibinfo {volume} {69}},\ \bibinfo {pages} {161301(R)}
  (\bibinfo {year} {2004})}\BibitemShut {NoStop}%
\bibitem [{\citenamefont {Young}\ \emph {et~al.}(2005)\citenamefont {Young},
  \citenamefont {Stevenson}, \citenamefont {Shields}, \citenamefont {Atkinson},
  \citenamefont {Cooper}, \citenamefont {Ritchie}, \citenamefont {Groom},
  \citenamefont {Tartakovskii},\ and\ \citenamefont {Skolnick}}]{Young_PRB05}%
  \BibitemOpen
  \bibfield  {author} {\bibinfo {author} {\bibfnamefont {R.~J.}\ \bibnamefont
  {Young}}, \bibinfo {author} {\bibfnamefont {R.~M.}\ \bibnamefont
  {Stevenson}}, \bibinfo {author} {\bibfnamefont {A.~J.}\ \bibnamefont
  {Shields}}, \bibinfo {author} {\bibfnamefont {P.}~\bibnamefont {Atkinson}},
  \bibinfo {author} {\bibfnamefont {K.}~\bibnamefont {Cooper}}, \bibinfo
  {author} {\bibfnamefont {D.~A.}\ \bibnamefont {Ritchie}}, \bibinfo {author}
  {\bibfnamefont {K.~M.}\ \bibnamefont {Groom}}, \bibinfo {author}
  {\bibfnamefont {A.~I.}\ \bibnamefont {Tartakovskii}}, \ and\ \bibinfo
  {author} {\bibfnamefont {M.~S.}\ \bibnamefont {Skolnick}},\ }\bibfield
  {title} {\enquote {\bibinfo {title} {Inversion of exciton level splitting in
  quantum dots},}\ }\href {\doibase 10.1103/PhysRevB.72.113305} {\bibfield
  {journal} {\bibinfo  {journal} {Phys. Rev. B}\ }\textbf {\bibinfo {volume}
  {72}},\ \bibinfo {pages} {113305} (\bibinfo {year} {2005})}\BibitemShut
  {NoStop}%
\bibitem [{\citenamefont {Akopian}\ \emph {et~al.}(2006)\citenamefont
  {Akopian}, \citenamefont {Lindner}, \citenamefont {Poem}, \citenamefont
  {Berlatzky}, \citenamefont {Avron}, \citenamefont {Gershoni}, \citenamefont
  {Gerardot},\ and\ \citenamefont {Petroff}}]{Akopian_PRL06}%
  \BibitemOpen
  \bibfield  {author} {\bibinfo {author} {\bibfnamefont {N.}~\bibnamefont
  {Akopian}}, \bibinfo {author} {\bibfnamefont {N.~H.}\ \bibnamefont
  {Lindner}}, \bibinfo {author} {\bibfnamefont {E.}~\bibnamefont {Poem}},
  \bibinfo {author} {\bibfnamefont {Y.}~\bibnamefont {Berlatzky}}, \bibinfo
  {author} {\bibfnamefont {J.}~\bibnamefont {Avron}}, \bibinfo {author}
  {\bibfnamefont {D.}~\bibnamefont {Gershoni}}, \bibinfo {author}
  {\bibfnamefont {B.~D.}\ \bibnamefont {Gerardot}}, \ and\ \bibinfo {author}
  {\bibfnamefont {P.~M.}\ \bibnamefont {Petroff}},\ }\bibfield  {title}
  {\enquote {\bibinfo {title} {Entangled photon pairs from semiconductor
  quantum dots},}\ }\href {\doibase 10.1103/PhysRevLett.96.130501} {\bibfield
  {journal} {\bibinfo  {journal} {Phys. Rev. Lett.}\ }\textbf {\bibinfo
  {volume} {96}},\ \bibinfo {pages} {130501} (\bibinfo {year}
  {2006})}\BibitemShut {NoStop}%
\bibitem [{\citenamefont {Stevenson}\ \emph {et~al.}(2006)\citenamefont
  {Stevenson}, \citenamefont {Young}, \citenamefont {Atkinson}, \citenamefont
  {Cooper}, \citenamefont {Ritchie},\ and\ \citenamefont
  {Shields}}]{Stevenson_Nat06}%
  \BibitemOpen
  \bibfield  {author} {\bibinfo {author} {\bibfnamefont {R.~M.}\ \bibnamefont
  {Stevenson}}, \bibinfo {author} {\bibfnamefont {R.~J.}\ \bibnamefont
  {Young}}, \bibinfo {author} {\bibfnamefont {P.}~\bibnamefont {Atkinson}},
  \bibinfo {author} {\bibfnamefont {K.}~\bibnamefont {Cooper}}, \bibinfo
  {author} {\bibfnamefont {D.~A.}\ \bibnamefont {Ritchie}}, \ and\ \bibinfo
  {author} {\bibfnamefont {A.~J.}\ \bibnamefont {Shields}},\ }\bibfield
  {title} {\enquote {\bibinfo {title} {A semiconductor source of triggered
  entangled photon pairs},}\ }\href@noop {} {\bibfield  {journal} {\bibinfo
  {journal} {Nature}\ }\textbf {\bibinfo {volume} {439}},\ \bibinfo {pages}
  {178--182} (\bibinfo {year} {2006})}\BibitemShut {NoStop}%
\bibitem [{\citenamefont {Pooley}\ \emph {et~al.}(2014)\citenamefont {Pooley},
  \citenamefont {Bennett}, \citenamefont {Stevenson}, \citenamefont {Shields},
  \citenamefont {Farrer},\ and\ \citenamefont {Ritchie}}]{Pooley_PRAppl14}%
  \BibitemOpen
  \bibfield  {author} {\bibinfo {author} {\bibfnamefont {M.~A.}\ \bibnamefont
  {Pooley}}, \bibinfo {author} {\bibfnamefont {A.~J.}\ \bibnamefont {Bennett}},
  \bibinfo {author} {\bibfnamefont {R.~M.}\ \bibnamefont {Stevenson}}, \bibinfo
  {author} {\bibfnamefont {A.~J.}\ \bibnamefont {Shields}}, \bibinfo {author}
  {\bibfnamefont {I.}~\bibnamefont {Farrer}}, \ and\ \bibinfo {author}
  {\bibfnamefont {D.~A.}\ \bibnamefont {Ritchie}},\ }\bibfield  {title}
  {\enquote {\bibinfo {title} {Energy-tunable quantum dot with minimal fine
  structure created by using simultaneous electric and magnetic fields},}\
  }\href {\doibase 10.1103/PhysRevApplied.1.024002} {\bibfield  {journal}
  {\bibinfo  {journal} {Phys. Rev. Applied}\ }\textbf {\bibinfo {volume} {1}},\
  \bibinfo {pages} {024002} (\bibinfo {year} {2014})}\BibitemShut {NoStop}%
\bibitem [{\citenamefont {Seidl}\ \emph {et~al.}(2006)\citenamefont {Seidl},
  \citenamefont {Kroner}, \citenamefont {H{\"o}gele}, \citenamefont {Karrai},
  \citenamefont {Warburton}, \citenamefont {Badolato},\ and\ \citenamefont
  {Petroff}}]{Seidl_APL06}%
  \BibitemOpen
  \bibfield  {author} {\bibinfo {author} {\bibfnamefont {S.}~\bibnamefont
  {Seidl}}, \bibinfo {author} {\bibfnamefont {M.}~\bibnamefont {Kroner}},
  \bibinfo {author} {\bibfnamefont {A.}~\bibnamefont {H{\"o}gele}}, \bibinfo
  {author} {\bibfnamefont {K.}~\bibnamefont {Karrai}}, \bibinfo {author}
  {\bibfnamefont {R.~J.}\ \bibnamefont {Warburton}}, \bibinfo {author}
  {\bibfnamefont {A.}~\bibnamefont {Badolato}}, \ and\ \bibinfo {author}
  {\bibfnamefont {P.~M.}\ \bibnamefont {Petroff}},\ }\bibfield  {title}
  {\enquote {\bibinfo {title} {Effect of uniaxial stress on excitons in a
  self-assembled quantum dot},}\ }\href {\doibase
  http://dx.doi.org/10.1063/1.2204843} {\bibfield  {journal} {\bibinfo
  {journal} {Appl. Phys. Lett.}\ }\textbf {\bibinfo {volume} {88}},\ \bibinfo
  {eid} {203113} (\bibinfo {year} {2006})}\BibitemShut {NoStop}%
\bibitem [{\citenamefont {Gerardot}\ \emph {et~al.}(2007)\citenamefont
  {Gerardot}, \citenamefont {Seidl}, \citenamefont {Dalgarno}, \citenamefont
  {Warburton}, \citenamefont {Granados}, \citenamefont {Garcia}, \citenamefont
  {Kowalik}, \citenamefont {Krebs}, \citenamefont {Karrai}, \citenamefont
  {Badolato},\ and\ \citenamefont {Petroff}}]{Gerardot_APL07}%
  \BibitemOpen
  \bibfield  {author} {\bibinfo {author} {\bibfnamefont {B.~D.}\ \bibnamefont
  {Gerardot}}, \bibinfo {author} {\bibfnamefont {S.}~\bibnamefont {Seidl}},
  \bibinfo {author} {\bibfnamefont {P.~A.}\ \bibnamefont {Dalgarno}}, \bibinfo
  {author} {\bibfnamefont {R.~J.}\ \bibnamefont {Warburton}}, \bibinfo {author}
  {\bibfnamefont {D.}~\bibnamefont {Granados}}, \bibinfo {author}
  {\bibfnamefont {J.~M.}\ \bibnamefont {Garcia}}, \bibinfo {author}
  {\bibfnamefont {K.}~\bibnamefont {Kowalik}}, \bibinfo {author} {\bibfnamefont
  {O.}~\bibnamefont {Krebs}}, \bibinfo {author} {\bibfnamefont
  {K.}~\bibnamefont {Karrai}}, \bibinfo {author} {\bibfnamefont
  {A.}~\bibnamefont {Badolato}}, \ and\ \bibinfo {author} {\bibfnamefont
  {P.~M.}\ \bibnamefont {Petroff}},\ }\bibfield  {title} {\enquote {\bibinfo
  {title} {Manipulating exciton fine structure in quantum dots with a lateral
  electric field},}\ }\href {\doibase http://dx.doi.org/10.1063/1.2431758}
  {\bibfield  {journal} {\bibinfo  {journal} {Appl. Phys. Lett.}\ }\textbf
  {\bibinfo {volume} {90}},\ \bibinfo {eid} {041101} (\bibinfo {year}
  {2007})}\BibitemShut {NoStop}%
\bibitem [{\citenamefont {Muller}\ \emph {et~al.}(2009)\citenamefont {Muller},
  \citenamefont {Fang}, \citenamefont {Lawall},\ and\ \citenamefont
  {Solomon}}]{Muller_PRL08}%
  \BibitemOpen
  \bibfield  {author} {\bibinfo {author} {\bibfnamefont {A.}~\bibnamefont
  {Muller}}, \bibinfo {author} {\bibfnamefont {W.}~\bibnamefont {Fang}},
  \bibinfo {author} {\bibfnamefont {J.}~\bibnamefont {Lawall}}, \ and\ \bibinfo
  {author} {\bibfnamefont {G.~S.}\ \bibnamefont {Solomon}},\ }\bibfield
  {title} {\enquote {\bibinfo {title} {Creating polarization-entangled photon
  pairs from a semiconductor quantum dot using the optical {Stark} effect},}\
  }\href {\doibase 10.1103/PhysRevLett.103.217402} {\bibfield  {journal}
  {\bibinfo  {journal} {Phys. Rev. Lett.}\ }\textbf {\bibinfo {volume} {103}},\
  \bibinfo {pages} {217402} (\bibinfo {year} {2009})}\BibitemShut {NoStop}%
\bibitem [{\citenamefont {Dousse}\ \emph {et~al.}(2010)\citenamefont {Dousse},
  \citenamefont {Suffczy\`nski}, \citenamefont {Beveratos}, \citenamefont
  {Krebs}, \citenamefont {Lema\^itre}, \citenamefont {Sagnes}, \citenamefont
  {Bloch}, \citenamefont {Voisin},\ and\ \citenamefont
  {Senellart}}]{Dousse_Nat10}%
  \BibitemOpen
  \bibfield  {author} {\bibinfo {author} {\bibfnamefont {A.}~\bibnamefont
  {Dousse}}, \bibinfo {author} {\bibfnamefont {J.}~\bibnamefont
  {Suffczy\`nski}}, \bibinfo {author} {\bibfnamefont {A.}~\bibnamefont
  {Beveratos}}, \bibinfo {author} {\bibfnamefont {O.}~\bibnamefont {Krebs}},
  \bibinfo {author} {\bibfnamefont {A.}~\bibnamefont {Lema\^itre}}, \bibinfo
  {author} {\bibfnamefont {I.}~\bibnamefont {Sagnes}}, \bibinfo {author}
  {\bibfnamefont {J.}~\bibnamefont {Bloch}}, \bibinfo {author} {\bibfnamefont
  {P.}~\bibnamefont {Voisin}}, \ and\ \bibinfo {author} {\bibfnamefont
  {P.}~\bibnamefont {Senellart}},\ }\bibfield  {title} {\enquote {\bibinfo
  {title} {Ultrabright source of entangled photon pairs},}\ }\href@noop {}
  {\bibfield  {journal} {\bibinfo  {journal} {Nature}\ }\textbf {\bibinfo
  {volume} {466}},\ \bibinfo {pages} {217--220} (\bibinfo {year}
  {2010})}\BibitemShut {NoStop}%
\bibitem [{\citenamefont {Ghali}\ \emph {et~al.}(2012)\citenamefont {Ghali},
  \citenamefont {Ohtani}, \citenamefont {Ohno},\ and\ \citenamefont
  {Ohno}}]{Ghali_NatComm12}%
  \BibitemOpen
  \bibfield  {author} {\bibinfo {author} {\bibfnamefont {M.}~\bibnamefont
  {Ghali}}, \bibinfo {author} {\bibfnamefont {K.}~\bibnamefont {Ohtani}},
  \bibinfo {author} {\bibfnamefont {Y.}~\bibnamefont {Ohno}}, \ and\ \bibinfo
  {author} {\bibfnamefont {H.}~\bibnamefont {Ohno}},\ }\bibfield  {title}
  {\enquote {\bibinfo {title} {Generation and control of polarization-entangled
  photons from {GaAs} island quantum dots by an electric field},}\ }\href
  {\doibase 10.1038/ncomms1657} {\bibfield  {journal} {\bibinfo  {journal}
  {Nat. Commun.}\ }\textbf {\bibinfo {volume} {3:661}} (\bibinfo {year}
  {2012}),\ 10.1038/ncomms1657}\BibitemShut {NoStop}%
\bibitem [{\citenamefont {Pooley}\ \emph {et~al.}(2013)\citenamefont {Pooley},
  \citenamefont {Bennett}, \citenamefont {Farrer}, \citenamefont {Ritchie},\
  and\ \citenamefont {Shields}}]{Pooley_APL13}%
  \BibitemOpen
  \bibfield  {author} {\bibinfo {author} {\bibfnamefont {M.~A.}\ \bibnamefont
  {Pooley}}, \bibinfo {author} {\bibfnamefont {A.~J.}\ \bibnamefont {Bennett}},
  \bibinfo {author} {\bibfnamefont {I.}~\bibnamefont {Farrer}}, \bibinfo
  {author} {\bibfnamefont {D.~A.}\ \bibnamefont {Ritchie}}, \ and\ \bibinfo
  {author} {\bibfnamefont {A.~J.}\ \bibnamefont {Shields}},\ }\bibfield
  {title} {\enquote {\bibinfo {title} {Engineering quantum dots for electrical
  control of the fine structure splitting},}\ }\href {\doibase
  http://dx.doi.org/10.1063/1.4813319} {\bibfield  {journal} {\bibinfo
  {journal} {Appl. Phys. Lett.}\ }\textbf {\bibinfo {volume} {103}},\ \bibinfo
  {eid} {031105} (\bibinfo {year} {2013})}\BibitemShut {NoStop}%
\bibitem [{\citenamefont {Trotta}\ \emph {et~al.}(2012)\citenamefont {Trotta},
  \citenamefont {Zallo}, \citenamefont {Ortix}, \citenamefont {Atkinson},
  \citenamefont {Plumhof}, \citenamefont {van~den Brink}, \citenamefont
  {Rastelli},\ and\ \citenamefont {Schmidt}}]{Trotta_PRL12}%
  \BibitemOpen
  \bibfield  {author} {\bibinfo {author} {\bibfnamefont {R.}~\bibnamefont
  {Trotta}}, \bibinfo {author} {\bibfnamefont {E.}~\bibnamefont {Zallo}},
  \bibinfo {author} {\bibfnamefont {C.}~\bibnamefont {Ortix}}, \bibinfo
  {author} {\bibfnamefont {P.}~\bibnamefont {Atkinson}}, \bibinfo {author}
  {\bibfnamefont {J.~D.}\ \bibnamefont {Plumhof}}, \bibinfo {author}
  {\bibfnamefont {J.}~\bibnamefont {van~den Brink}}, \bibinfo {author}
  {\bibfnamefont {A.}~\bibnamefont {Rastelli}}, \ and\ \bibinfo {author}
  {\bibfnamefont {O.~G.}\ \bibnamefont {Schmidt}},\ }\bibfield  {title}
  {\enquote {\bibinfo {title} {Universal recovery of the energy-level
  degeneracy of bright excitons in {InGaAs} quantum dots without a structure
  symmetry},}\ }\href {\doibase 10.1103/PhysRevLett.109.147401} {\bibfield
  {journal} {\bibinfo  {journal} {Phys. Rev. Lett.}\ }\textbf {\bibinfo
  {volume} {109}},\ \bibinfo {pages} {147401} (\bibinfo {year}
  {2012})}\BibitemShut {NoStop}%
\bibitem [{\citenamefont {Trotta}\ \emph {et~al.}()\citenamefont {Trotta},
  \citenamefont {Wildmann}, \citenamefont {Zallo}, \citenamefont {Schmidt},\
  and\ \citenamefont {Rastelli}}]{Trotta_arxiv14}%
  \BibitemOpen
  \bibfield  {author} {\bibinfo {author} {\bibfnamefont {R.}~\bibnamefont
  {Trotta}}, \bibinfo {author} {\bibfnamefont {J.~S.}\ \bibnamefont
  {Wildmann}}, \bibinfo {author} {\bibfnamefont {E.}~\bibnamefont {Zallo}},
  \bibinfo {author} {\bibfnamefont {O.~G.}\ \bibnamefont {Schmidt}}, \ and\
  \bibinfo {author} {\bibfnamefont {A.}~\bibnamefont {Rastelli}},\ }\href@noop
  {} {\enquote {\bibinfo {title} {Highly entangled photons from hybrid
  piezoelectric-semiconductor quantum dot devices},}\ }\bibinfo {howpublished}
  {arXiv:1403.0225}\BibitemShut {NoStop}%
\bibitem [{\citenamefont {Singh}\ and\ \citenamefont
  {Bester}(2009)}]{Sing_PRL09}%
  \BibitemOpen
  \bibfield  {author} {\bibinfo {author} {\bibfnamefont {R.}~\bibnamefont
  {Singh}}\ and\ \bibinfo {author} {\bibfnamefont {G.}~\bibnamefont {Bester}},\
  }\bibfield  {title} {\enquote {\bibinfo {title} {Nanowire quantum dots as an
  ideal source of entangled photon pairs},}\ }\href {\doibase
  10.1103/PhysRevLett.103.063601} {\bibfield  {journal} {\bibinfo  {journal}
  {Phys. Rev. Lett.}\ }\textbf {\bibinfo {volume} {103}},\ \bibinfo {pages}
  {063601} (\bibinfo {year} {2009})}\BibitemShut {NoStop}%
\bibitem [{\citenamefont {Schliwa}\ \emph {et~al.}(2009)\citenamefont
  {Schliwa}, \citenamefont {Winkelnkemper}, \citenamefont {Lochmann},
  \citenamefont {Stock},\ and\ \citenamefont {Bimberg}}]{Schliwa_PRB09}%
  \BibitemOpen
  \bibfield  {author} {\bibinfo {author} {\bibfnamefont {A.}~\bibnamefont
  {Schliwa}}, \bibinfo {author} {\bibfnamefont {M.}~\bibnamefont
  {Winkelnkemper}}, \bibinfo {author} {\bibfnamefont {A.}~\bibnamefont
  {Lochmann}}, \bibinfo {author} {\bibfnamefont {E.}~\bibnamefont {Stock}}, \
  and\ \bibinfo {author} {\bibfnamefont {D.}~\bibnamefont {Bimberg}},\
  }\bibfield  {title} {\enquote {\bibinfo {title} {{In(Ga)As/GaAs} quantum dots
  grown on a (111) surface as ideal sources of entangled photon pairs},}\
  }\href {\doibase 10.1103/PhysRevB.80.161307} {\bibfield  {journal} {\bibinfo
  {journal} {Phys. Rev. B}\ }\textbf {\bibinfo {volume} {80}},\ \bibinfo
  {pages} {161307} (\bibinfo {year} {2009})}\BibitemShut {NoStop}%
\bibitem [{\citenamefont {Sugiyama}\ \emph {et~al.}(1995)\citenamefont
  {Sugiyama}, \citenamefont {Sakuma}, \citenamefont {Muto},\ and\ \citenamefont
  {Yokoyama}}]{Sugiyama_APL95}%
  \BibitemOpen
  \bibfield  {author} {\bibinfo {author} {\bibfnamefont {Y.}~\bibnamefont
  {Sugiyama}}, \bibinfo {author} {\bibfnamefont {Y.}~\bibnamefont {Sakuma}},
  \bibinfo {author} {\bibfnamefont {S.}~\bibnamefont {Muto}}, \ and\ \bibinfo
  {author} {\bibfnamefont {N.}~\bibnamefont {Yokoyama}},\ }\bibfield  {title}
  {\enquote {\bibinfo {title} {Novel {InGaAs/GaAs} quantum dot structures
  formed in tetrahedral-shaped recesses on (111){B} {GaAs} substrate using
  metalorganic vapor phase epitaxy},}\ }\href {\doibase 10.1063/1.114685}
  {\bibfield  {journal} {\bibinfo  {journal} {Appl. Phys. Lett.}\ }\textbf
  {\bibinfo {volume} {67}},\ \bibinfo {pages} {256--258} (\bibinfo {year}
  {1995})}\BibitemShut {NoStop}%
\bibitem [{\citenamefont {Hartmann}\ \emph {et~al.}(1998)\citenamefont
  {Hartmann}, \citenamefont {Ducommun}, \citenamefont {Loubies}, \citenamefont
  {Leifer},\ and\ \citenamefont {Kapon}}]{Hartmann_APL98}%
  \BibitemOpen
  \bibfield  {author} {\bibinfo {author} {\bibfnamefont {A.}~\bibnamefont
  {Hartmann}}, \bibinfo {author} {\bibfnamefont {Y.}~\bibnamefont {Ducommun}},
  \bibinfo {author} {\bibfnamefont {L.}~\bibnamefont {Loubies}}, \bibinfo
  {author} {\bibfnamefont {K.}~\bibnamefont {Leifer}}, \ and\ \bibinfo {author}
  {\bibfnamefont {E.}~\bibnamefont {Kapon}},\ }\bibfield  {title} {\enquote
  {\bibinfo {title} {Structure and photoluminescence of single {AlGaAs/GaAs}
  quantum dots grown in inverted tetrahedral pyramids},}\ }\href {\doibase
  http://dx.doi.org/10.1063/1.121810} {\bibfield  {journal} {\bibinfo
  {journal} {Appl. Phys. Lett.}\ }\textbf {\bibinfo {volume} {73}},\ \bibinfo
  {pages} {2322--2324} (\bibinfo {year} {1998})}\BibitemShut {NoStop}%
\bibitem [{\citenamefont {Mereni}\ \emph {et~al.}(2009)\citenamefont {Mereni},
  \citenamefont {Dimastrodonato}, \citenamefont {Young},\ and\ \citenamefont
  {Pelucchi}}]{Mereni_APL09}%
  \BibitemOpen
  \bibfield  {author} {\bibinfo {author} {\bibfnamefont {L.~O.}\ \bibnamefont
  {Mereni}}, \bibinfo {author} {\bibfnamefont {V.}~\bibnamefont
  {Dimastrodonato}}, \bibinfo {author} {\bibfnamefont {R.~J.}\ \bibnamefont
  {Young}}, \ and\ \bibinfo {author} {\bibfnamefont {E.}~\bibnamefont
  {Pelucchi}},\ }\bibfield  {title} {\enquote {\bibinfo {title} {A
  site-controlled quantum dot system offering both high uniformity and spectral
  purity},}\ }\href {\doibase http://dx.doi.org/10.1063/1.3147213} {\bibfield
  {journal} {\bibinfo  {journal} {Appl. Phys. Lett.}\ }\textbf {\bibinfo
  {volume} {94}},\ \bibinfo {eid} {223121} (\bibinfo {year}
  {2009})}\BibitemShut {NoStop}%
\bibitem [{\citenamefont {Stock}\ \emph {et~al.}(2010)\citenamefont {Stock},
  \citenamefont {Warming}, \citenamefont {Ostapenko}, \citenamefont {Rodt},
  \citenamefont {Schliwa}, \citenamefont {T\"{o}fflinger}, \citenamefont
  {Lochmann}, \citenamefont {Toropov}, \citenamefont {Moshchenko},
  \citenamefont {Dmitriev}, \citenamefont {Haisler},\ and\ \citenamefont
  {Bimberg}}]{Stock_APL10}%
  \BibitemOpen
  \bibfield  {author} {\bibinfo {author} {\bibfnamefont {E.}~\bibnamefont
  {Stock}}, \bibinfo {author} {\bibfnamefont {T.}~\bibnamefont {Warming}},
  \bibinfo {author} {\bibfnamefont {I.}~\bibnamefont {Ostapenko}}, \bibinfo
  {author} {\bibfnamefont {S.}~\bibnamefont {Rodt}}, \bibinfo {author}
  {\bibfnamefont {A.}~\bibnamefont {Schliwa}}, \bibinfo {author} {\bibfnamefont
  {J.~A.}\ \bibnamefont {T\"{o}fflinger}}, \bibinfo {author} {\bibfnamefont
  {A.}~\bibnamefont {Lochmann}}, \bibinfo {author} {\bibfnamefont {A.~I.}\
  \bibnamefont {Toropov}}, \bibinfo {author} {\bibfnamefont {S.~A.}\
  \bibnamefont {Moshchenko}}, \bibinfo {author} {\bibfnamefont {D.~V.}\
  \bibnamefont {Dmitriev}}, \bibinfo {author} {\bibfnamefont {V.~A.}\
  \bibnamefont {Haisler}}, \ and\ \bibinfo {author} {\bibfnamefont
  {D.}~\bibnamefont {Bimberg}},\ }\bibfield  {title} {\enquote {\bibinfo
  {title} {Single-photon emission from {InGaAs} quantum dots grown on (111)
  {GaAs}},}\ }\href {\doibase 10.1063/1.3337097} {\bibfield  {journal}
  {\bibinfo  {journal} {Appl. Phys. Lett.}\ }\textbf {\bibinfo {volume} {96}},\
  \bibinfo {eid} {093112} (\bibinfo {year} {2010})}\BibitemShut {NoStop}%
\bibitem [{\citenamefont {Mano}\ \emph {et~al.}(2010)\citenamefont {Mano},
  \citenamefont {Abbarchi}, \citenamefont {Kuroda}, \citenamefont {McSkimming},
  \citenamefont {Ohtake}, \citenamefont {Mitsuishi},\ and\ \citenamefont
  {Sakoda}}]{Mano2010}%
  \BibitemOpen
  \bibfield  {author} {\bibinfo {author} {\bibfnamefont {T.}~\bibnamefont
  {Mano}}, \bibinfo {author} {\bibfnamefont {M.}~\bibnamefont {Abbarchi}},
  \bibinfo {author} {\bibfnamefont {T.}~\bibnamefont {Kuroda}}, \bibinfo
  {author} {\bibfnamefont {B.}~\bibnamefont {McSkimming}}, \bibinfo {author}
  {\bibfnamefont {A.}~\bibnamefont {Ohtake}}, \bibinfo {author} {\bibfnamefont
  {K.}~\bibnamefont {Mitsuishi}}, \ and\ \bibinfo {author} {\bibfnamefont
  {K.}~\bibnamefont {Sakoda}},\ }\bibfield  {title} {\enquote {\bibinfo {title}
  {Self-assembly of symmetric {GaAs} quantum dots on {(111)A} substrates:
  Suppression of fine-structure splitting},}\ }\href
  {http://stacks.iop.org/1882-0786/3/i=6/a=065203} {\bibfield  {journal}
  {\bibinfo  {journal} {Appl. Phys. Express}\ }\textbf {\bibinfo {volume}
  {3}},\ \bibinfo {pages} {065203} (\bibinfo {year} {2010})}\BibitemShut
  {NoStop}%
\bibitem [{\citenamefont {Jo}\ \emph {et~al.}(2012)\citenamefont {Jo},
  \citenamefont {Mano}, \citenamefont {Abbarchi}, \citenamefont {Kuroda},
  \citenamefont {Sakuma},\ and\ \citenamefont {Sakoda}}]{Jo_CGD12}%
  \BibitemOpen
  \bibfield  {author} {\bibinfo {author} {\bibfnamefont {M.}~\bibnamefont
  {Jo}}, \bibinfo {author} {\bibfnamefont {T.}~\bibnamefont {Mano}}, \bibinfo
  {author} {\bibfnamefont {M.}~\bibnamefont {Abbarchi}}, \bibinfo {author}
  {\bibfnamefont {T.}~\bibnamefont {Kuroda}}, \bibinfo {author} {\bibfnamefont
  {Y.}~\bibnamefont {Sakuma}}, \ and\ \bibinfo {author} {\bibfnamefont
  {K.}~\bibnamefont {Sakoda}},\ }\bibfield  {title} {\enquote {\bibinfo {title}
  {Self-limiting growth of hexagonal and triangular quantum dots on
  (111){A}},}\ }\href {\doibase 10.1021/cg201513m} {\bibfield  {journal}
  {\bibinfo  {journal} {Cryst. Growth Des.}\ }\textbf {\bibinfo {volume}
  {12}},\ \bibinfo {pages} {1411--1415} (\bibinfo {year} {2012})}\BibitemShut
  {NoStop}%
\bibitem [{\citenamefont {Karlsson}\ \emph {et~al.}(2010)\citenamefont
  {Karlsson}, \citenamefont {Dupertuis}, \citenamefont {Oberli}, \citenamefont
  {Pelucchi}, \citenamefont {Rudra}, \citenamefont {Holtz},\ and\ \citenamefont
  {Kapon}}]{Karlsson_PRB10}%
  \BibitemOpen
  \bibfield  {author} {\bibinfo {author} {\bibfnamefont {K.~F.}\ \bibnamefont
  {Karlsson}}, \bibinfo {author} {\bibfnamefont {M.~A.}\ \bibnamefont
  {Dupertuis}}, \bibinfo {author} {\bibfnamefont {D.~Y.}\ \bibnamefont
  {Oberli}}, \bibinfo {author} {\bibfnamefont {E.}~\bibnamefont {Pelucchi}},
  \bibinfo {author} {\bibfnamefont {A.}~\bibnamefont {Rudra}}, \bibinfo
  {author} {\bibfnamefont {P.~O.}\ \bibnamefont {Holtz}}, \ and\ \bibinfo
  {author} {\bibfnamefont {E.}~\bibnamefont {Kapon}},\ }\bibfield  {title}
  {\enquote {\bibinfo {title} {Fine structure of exciton complexes in
  high-symmetry quantum dots: {E}ffects of symmetry breaking and symmetry
  elevation},}\ }\href {\doibase 10.1103/PhysRevB.81.161307} {\bibfield
  {journal} {\bibinfo  {journal} {Phys. Rev. B}\ }\textbf {\bibinfo {volume}
  {81}},\ \bibinfo {pages} {161307} (\bibinfo {year} {2010})}\BibitemShut
  {NoStop}%
\bibitem [{\citenamefont {Mereni}\ \emph {et~al.}(2012)\citenamefont {Mereni},
  \citenamefont {Marquardt}, \citenamefont {Juska}, \citenamefont
  {Dimastrodonato}, \citenamefont {O'Reilly},\ and\ \citenamefont
  {Pelucchi}}]{Mereni_PRB10}%
  \BibitemOpen
  \bibfield  {author} {\bibinfo {author} {\bibfnamefont {L.~O.}\ \bibnamefont
  {Mereni}}, \bibinfo {author} {\bibfnamefont {O.}~\bibnamefont {Marquardt}},
  \bibinfo {author} {\bibfnamefont {G.}~\bibnamefont {Juska}}, \bibinfo
  {author} {\bibfnamefont {V.}~\bibnamefont {Dimastrodonato}}, \bibinfo
  {author} {\bibfnamefont {E.~P.}\ \bibnamefont {O'Reilly}}, \ and\ \bibinfo
  {author} {\bibfnamefont {E.}~\bibnamefont {Pelucchi}},\ }\bibfield  {title}
  {\enquote {\bibinfo {title} {Fine-structure splitting in large-pitch
  pyramidal quantum dots},}\ }\href {\doibase 10.1103/PhysRevB.85.155453}
  {\bibfield  {journal} {\bibinfo  {journal} {Phys. Rev. B}\ }\textbf {\bibinfo
  {volume} {85}},\ \bibinfo {pages} {155453} (\bibinfo {year}
  {2012})}\BibitemShut {NoStop}%
\bibitem [{\citenamefont {Juska}\ \emph {et~al.}(2013)\citenamefont {Juska},
  \citenamefont {Dimastrodonato}, \citenamefont {Mereni}, \citenamefont
  {Gocalinska},\ and\ \citenamefont {Pelucchi}}]{Juska2013}%
  \BibitemOpen
  \bibfield  {author} {\bibinfo {author} {\bibfnamefont {G.}~\bibnamefont
  {Juska}}, \bibinfo {author} {\bibfnamefont {V.}~\bibnamefont
  {Dimastrodonato}}, \bibinfo {author} {\bibfnamefont {L.~O.}\ \bibnamefont
  {Mereni}}, \bibinfo {author} {\bibfnamefont {A.}~\bibnamefont {Gocalinska}},
  \ and\ \bibinfo {author} {\bibfnamefont {E.}~\bibnamefont {Pelucchi}},\
  }\bibfield  {title} {\enquote {\bibinfo {title} {Towards quantum-dot arrays
  of entangled photon emitters},}\ }\href {\doibase 10.1038/NPHOTON.2013.128}
  {\bibfield  {journal} {\bibinfo  {journal} {Nature Photon.}\ }\textbf
  {\bibinfo {volume} {7}},\ \bibinfo {pages} {527} (\bibinfo {year}
  {2013})}\BibitemShut {NoStop}%
\bibitem [{\citenamefont {Kuroda}\ \emph {et~al.}(2013)\citenamefont {Kuroda},
  \citenamefont {Mano}, \citenamefont {Ha}, \citenamefont {Nakajima},
  \citenamefont {Kumano}, \citenamefont {Urbaszek}, \citenamefont {Jo},
  \citenamefont {Abbarchi}, \citenamefont {Sakuma}, \citenamefont {Sakoda},
  \citenamefont {Suemune}, \citenamefont {Marie},\ and\ \citenamefont
  {Amand}}]{Kuroda2013}%
  \BibitemOpen
  \bibfield  {author} {\bibinfo {author} {\bibfnamefont {T.}~\bibnamefont
  {Kuroda}}, \bibinfo {author} {\bibfnamefont {T.}~\bibnamefont {Mano}},
  \bibinfo {author} {\bibfnamefont {N.}~\bibnamefont {Ha}}, \bibinfo {author}
  {\bibfnamefont {H.}~\bibnamefont {Nakajima}}, \bibinfo {author}
  {\bibfnamefont {H.}~\bibnamefont {Kumano}}, \bibinfo {author} {\bibfnamefont
  {B.}~\bibnamefont {Urbaszek}}, \bibinfo {author} {\bibfnamefont
  {M.}~\bibnamefont {Jo}}, \bibinfo {author} {\bibfnamefont {M.}~\bibnamefont
  {Abbarchi}}, \bibinfo {author} {\bibfnamefont {Y.}~\bibnamefont {Sakuma}},
  \bibinfo {author} {\bibfnamefont {K.}~\bibnamefont {Sakoda}}, \bibinfo
  {author} {\bibfnamefont {I.}~\bibnamefont {Suemune}}, \bibinfo {author}
  {\bibfnamefont {X.}~\bibnamefont {Marie}}, \ and\ \bibinfo {author}
  {\bibfnamefont {T.}~\bibnamefont {Amand}},\ }\bibfield  {title} {\enquote
  {\bibinfo {title} {Symmetric quantum dots as efficient sources of highly
  entangled photons: {V}iolation of {B}ell's inequality without spectral and
  temporal filtering},}\ }\href {\doibase 10.1103/PhysRevB.88.041306}
  {\bibfield  {journal} {\bibinfo  {journal} {Phys. Phys. B}\ }\textbf
  {\bibinfo {volume} {88}},\ \bibinfo {pages} {041306(R)} (\bibinfo {year}
  {2013})}\BibitemShut {NoStop}%
\bibitem [{\citenamefont {Ha}\ \emph {et~al.}(2014)\citenamefont {Ha},
  \citenamefont {Liu}, \citenamefont {Mano}, \citenamefont {Kuroda},
  \citenamefont {Mitsuishi}, \citenamefont {Castellano}, \citenamefont
  {Sanguinetti}, \citenamefont {Noda}, \citenamefont {Sakuma},\ and\
  \citenamefont {Sakoda}}]{Ha_APL14}%
  \BibitemOpen
  \bibfield  {author} {\bibinfo {author} {\bibfnamefont {N.}~\bibnamefont
  {Ha}}, \bibinfo {author} {\bibfnamefont {X.}~\bibnamefont {Liu}}, \bibinfo
  {author} {\bibfnamefont {T.}~\bibnamefont {Mano}}, \bibinfo {author}
  {\bibfnamefont {T.}~\bibnamefont {Kuroda}}, \bibinfo {author} {\bibfnamefont
  {K.}~\bibnamefont {Mitsuishi}}, \bibinfo {author} {\bibfnamefont
  {A.}~\bibnamefont {Castellano}}, \bibinfo {author} {\bibfnamefont
  {S.}~\bibnamefont {Sanguinetti}}, \bibinfo {author} {\bibfnamefont
  {T.}~\bibnamefont {Noda}}, \bibinfo {author} {\bibfnamefont {Y.}~\bibnamefont
  {Sakuma}}, \ and\ \bibinfo {author} {\bibfnamefont {K.}~\bibnamefont
  {Sakoda}},\ }\bibfield  {title} {\enquote {\bibinfo {title} {Droplet
  epitaxial growth of highly symmetric quantum dots emitting at
  telecommunication wavelengths on {InP(111)A}},}\ }\href {\doibase
  http://dx.doi.org/10.1063/1.4870839} {\bibfield  {journal} {\bibinfo
  {journal} {Appl. Phys. Lett.}\ }\textbf {\bibinfo {volume} {104}},\ \bibinfo
  {eid} {143106} (\bibinfo {year} {2014})}\BibitemShut {NoStop}%
\bibitem [{\citenamefont {Cade}\ \emph {et~al.}(2006)\citenamefont {Cade},
  \citenamefont {Gotoh}, \citenamefont {Kamada}, \citenamefont {Nakano},\ and\
  \citenamefont {Okamoto}}]{Cade_PRB06}%
  \BibitemOpen
  \bibfield  {author} {\bibinfo {author} {\bibfnamefont {N.~I.}\ \bibnamefont
  {Cade}}, \bibinfo {author} {\bibfnamefont {H.}~\bibnamefont {Gotoh}},
  \bibinfo {author} {\bibfnamefont {H.}~\bibnamefont {Kamada}}, \bibinfo
  {author} {\bibfnamefont {H.}~\bibnamefont {Nakano}}, \ and\ \bibinfo {author}
  {\bibfnamefont {H.}~\bibnamefont {Okamoto}},\ }\bibfield  {title} {\enquote
  {\bibinfo {title} {Fine structure and magneto-optics of exciton, trion, and
  charged biexciton states in single {InAs} quantum dots emitting at 1.3
  $\mu${}m},}\ }\href {\doibase 10.1103/PhysRevB.73.115322} {\bibfield
  {journal} {\bibinfo  {journal} {Phys. Rev. B}\ }\textbf {\bibinfo {volume}
  {73}},\ \bibinfo {pages} {115322} (\bibinfo {year} {2006})}\BibitemShut
  {NoStop}%
\bibitem [{\citenamefont {Chauvin}\ \emph {et~al.}(2006)\citenamefont
  {Chauvin}, \citenamefont {Salem}, \citenamefont {Bremond}, \citenamefont
  {Guillot}, \citenamefont {Bru-Chevallier},\ and\ \citenamefont
  {Gendry}}]{Chauvin2006}%
  \BibitemOpen
  \bibfield  {author} {\bibinfo {author} {\bibfnamefont {N.}~\bibnamefont
  {Chauvin}}, \bibinfo {author} {\bibfnamefont {B.}~\bibnamefont {Salem}},
  \bibinfo {author} {\bibfnamefont {G.}~\bibnamefont {Bremond}}, \bibinfo
  {author} {\bibfnamefont {G.}~\bibnamefont {Guillot}}, \bibinfo {author}
  {\bibfnamefont {C.}~\bibnamefont {Bru-Chevallier}}, \ and\ \bibinfo {author}
  {\bibfnamefont {M.}~\bibnamefont {Gendry}},\ }\bibfield  {title} {\enquote
  {\bibinfo {title} {Size and shape effects on excitons and biexcitons in
  single {InAs/InP} quantum dots},}\ }\href {\doibase
  http://dx.doi.org/10.1063/1.2353896} {\bibfield  {journal} {\bibinfo
  {journal} {J. Appl. Phys.}\ }\textbf {\bibinfo {volume} {100}},\ \bibinfo
  {eid} {073702} (\bibinfo {year} {2006})}\BibitemShut {NoStop}%
\bibitem [{\citenamefont {Sapienza}\ \emph {et~al.}(2013)\citenamefont
  {Sapienza}, \citenamefont {Malein}, \citenamefont {Kuklewicz}, \citenamefont
  {Kremer}, \citenamefont {Srinivasan}, \citenamefont {Griffiths},
  \citenamefont {Clarke}, \citenamefont {Gong}, \citenamefont {Warburton},\
  and\ \citenamefont {Gerardot}}]{Sapienza2013}%
  \BibitemOpen
  \bibfield  {author} {\bibinfo {author} {\bibfnamefont {L.}~\bibnamefont
  {Sapienza}}, \bibinfo {author} {\bibfnamefont {R.~N.~E.}\ \bibnamefont
  {Malein}}, \bibinfo {author} {\bibfnamefont {C.~E.}\ \bibnamefont
  {Kuklewicz}}, \bibinfo {author} {\bibfnamefont {P.~E.}\ \bibnamefont
  {Kremer}}, \bibinfo {author} {\bibfnamefont {K.}~\bibnamefont {Srinivasan}},
  \bibinfo {author} {\bibfnamefont {A.}~\bibnamefont {Griffiths}}, \bibinfo
  {author} {\bibfnamefont {E.}~\bibnamefont {Clarke}}, \bibinfo {author}
  {\bibfnamefont {M.}~\bibnamefont {Gong}}, \bibinfo {author} {\bibfnamefont
  {R.~J.}\ \bibnamefont {Warburton}}, \ and\ \bibinfo {author} {\bibfnamefont
  {B.~D.}\ \bibnamefont {Gerardot}},\ }\bibfield  {title} {\enquote {\bibinfo
  {title} {Exciton fine-structure splitting of telecom-wavelength single
  quantum dots: Statistics and external strain tuning},}\ }\href {\doibase
  10.1103/PhysRevB.88.155330} {\bibfield  {journal} {\bibinfo  {journal} {Phys.
  Rev. B}\ }\textbf {\bibinfo {volume} {88}},\ \bibinfo {pages} {155330}
  (\bibinfo {year} {2013})}\BibitemShut {NoStop}%
\bibitem [{\citenamefont {Ward}\ \emph {et~al.}(2014)\citenamefont {Ward},
  \citenamefont {Dean}, \citenamefont {Stevenson}, \citenamefont {Bennett},
  \citenamefont {Ellis}, \citenamefont {Cooper}, \citenamefont {Farrer},
  \citenamefont {Nicoll}, \citenamefont {Ritchie},\ and\ \citenamefont
  {Shields}}]{Ward_natcomm14}%
  \BibitemOpen
  \bibfield  {author} {\bibinfo {author} {\bibfnamefont {M.~B.}\ \bibnamefont
  {Ward}}, \bibinfo {author} {\bibfnamefont {M.~C.}\ \bibnamefont {Dean}},
  \bibinfo {author} {\bibfnamefont {R.~M.}\ \bibnamefont {Stevenson}}, \bibinfo
  {author} {\bibfnamefont {A.~J.}\ \bibnamefont {Bennett}}, \bibinfo {author}
  {\bibfnamefont {D.~J.~P.}\ \bibnamefont {Ellis}}, \bibinfo {author}
  {\bibfnamefont {K.}~\bibnamefont {Cooper}}, \bibinfo {author} {\bibfnamefont
  {I.}~\bibnamefont {Farrer}}, \bibinfo {author} {\bibfnamefont {C.~A.}\
  \bibnamefont {Nicoll}}, \bibinfo {author} {\bibfnamefont {D.~A.}\
  \bibnamefont {Ritchie}}, \ and\ \bibinfo {author} {\bibfnamefont {A.~J.}\
  \bibnamefont {Shields}},\ }\bibfield  {title} {\enquote {\bibinfo {title}
  {Coherent dynamics of a telecom-wavelength entangled photon source},}\ }\href
  {\doibase 10.1038/ncomms4316} {\bibfield  {journal} {\bibinfo  {journal}
  {Nat. Commun.}\ }\textbf {\bibinfo {volume} {5:3316}} (\bibinfo {year}
  {2014}),\ 10.1038/ncomms4316}\BibitemShut {NoStop}%
\bibitem [{\citenamefont {Koguchi}, \citenamefont {Takahashi},\ and\
  \citenamefont {Chikyow}(1991)}]{Koguchi1991}%
  \BibitemOpen
  \bibfield  {author} {\bibinfo {author} {\bibfnamefont {N.}~\bibnamefont
  {Koguchi}}, \bibinfo {author} {\bibfnamefont {S.}~\bibnamefont {Takahashi}},
  \ and\ \bibinfo {author} {\bibfnamefont {T.}~\bibnamefont {Chikyow}},\
  }\bibfield  {title} {\enquote {\bibinfo {title} {New {MBE} growth method for
  {InSb} quantum well boxes},}\ }\href {\doibase 10.1016/0022-0248(91)91064-H}
  {\bibfield  {journal} {\bibinfo  {journal} {J. Cryst. Growth}\ }\textbf
  {\bibinfo {volume} {111}},\ \bibinfo {pages} {688--692} (\bibinfo {year}
  {1991})}\BibitemShut {NoStop}%
\bibitem [{\citenamefont {Mano}\ \emph {et~al.}(2009)\citenamefont {Mano},
  \citenamefont {Abbarchi}, \citenamefont {Kuroda}, \citenamefont
  {Mastrandrea}, \citenamefont {Vinattieri}, \citenamefont {Sanguinetti},
  \citenamefont {Sakoda},\ and\ \citenamefont
  {Gurioli}}]{Mano_Nanotechnology09}%
  \BibitemOpen
  \bibfield  {author} {\bibinfo {author} {\bibfnamefont {T.}~\bibnamefont
  {Mano}}, \bibinfo {author} {\bibfnamefont {M.}~\bibnamefont {Abbarchi}},
  \bibinfo {author} {\bibfnamefont {T.}~\bibnamefont {Kuroda}}, \bibinfo
  {author} {\bibfnamefont {C.~A.}\ \bibnamefont {Mastrandrea}}, \bibinfo
  {author} {\bibfnamefont {A.}~\bibnamefont {Vinattieri}}, \bibinfo {author}
  {\bibfnamefont {S.}~\bibnamefont {Sanguinetti}}, \bibinfo {author}
  {\bibfnamefont {K.}~\bibnamefont {Sakoda}}, \ and\ \bibinfo {author}
  {\bibfnamefont {M.}~\bibnamefont {Gurioli}},\ }\bibfield  {title} {\enquote
  {\bibinfo {title} {Ultra-narrow emission from single {GaAs} self-assembled
  quantum dots grown by droplet epitaxy},}\ }\href
  {http://stacks.iop.org/0957-4484/20/i=39/a=395601} {\bibfield  {journal}
  {\bibinfo  {journal} {Nanotechnology}\ }\textbf {\bibinfo {volume} {20}},\
  \bibinfo {pages} {395601} (\bibinfo {year} {2009})}\BibitemShut {NoStop}%
\bibitem [{\citenamefont {Sallen}\ \emph {et~al.}(2011)\citenamefont {Sallen},
  \citenamefont {Urbaszek}, \citenamefont {Glazov}, \citenamefont {Ivchenko},
  \citenamefont {Kuroda}, \citenamefont {Mano}, \citenamefont {Kunz},
  \citenamefont {Abbarchi}, \citenamefont {Sakoda}, \citenamefont {Lagarde},
  \citenamefont {Balocchi}, \citenamefont {Marie},\ and\ \citenamefont
  {Amand}}]{Sallen_PRL11}%
  \BibitemOpen
  \bibfield  {author} {\bibinfo {author} {\bibfnamefont {G.}~\bibnamefont
  {Sallen}}, \bibinfo {author} {\bibfnamefont {B.}~\bibnamefont {Urbaszek}},
  \bibinfo {author} {\bibfnamefont {M.~M.}\ \bibnamefont {Glazov}}, \bibinfo
  {author} {\bibfnamefont {E.~L.}\ \bibnamefont {Ivchenko}}, \bibinfo {author}
  {\bibfnamefont {T.}~\bibnamefont {Kuroda}}, \bibinfo {author} {\bibfnamefont
  {T.}~\bibnamefont {Mano}}, \bibinfo {author} {\bibfnamefont {S.}~\bibnamefont
  {Kunz}}, \bibinfo {author} {\bibfnamefont {M.}~\bibnamefont {Abbarchi}},
  \bibinfo {author} {\bibfnamefont {K.}~\bibnamefont {Sakoda}}, \bibinfo
  {author} {\bibfnamefont {D.}~\bibnamefont {Lagarde}}, \bibinfo {author}
  {\bibfnamefont {A.}~\bibnamefont {Balocchi}}, \bibinfo {author}
  {\bibfnamefont {X.}~\bibnamefont {Marie}}, \ and\ \bibinfo {author}
  {\bibfnamefont {T.}~\bibnamefont {Amand}},\ }\bibfield  {title} {\enquote
  {\bibinfo {title} {Dark-bright mixing of interband transitions in symmetric
  semiconductor quantum dots},}\ }\href {\doibase
  10.1103/PhysRevLett.107.166604} {\bibfield  {journal} {\bibinfo  {journal}
  {Phys. Rev. Lett.}\ }\textbf {\bibinfo {volume} {107}},\ \bibinfo {pages}
  {166604} (\bibinfo {year} {2011})}\BibitemShut {NoStop}%
\bibitem [{\citenamefont {Sallen}\ \emph {et~al.}(2014)\citenamefont {Sallen},
  \citenamefont {Kunz}, \citenamefont {Amand}, \citenamefont {Bouet},
  \citenamefont {Kuroda}, \citenamefont {Mano}, \citenamefont {Paget},
  \citenamefont {Krebs}, \citenamefont {Marie}, \citenamefont {Sakoda},\ and\
  \citenamefont {Urbaszek}}]{Sallen_NatComm14}%
  \BibitemOpen
  \bibfield  {author} {\bibinfo {author} {\bibfnamefont {G.}~\bibnamefont
  {Sallen}}, \bibinfo {author} {\bibfnamefont {S.}~\bibnamefont {Kunz}},
  \bibinfo {author} {\bibfnamefont {T.}~\bibnamefont {Amand}}, \bibinfo
  {author} {\bibfnamefont {L.}~\bibnamefont {Bouet}}, \bibinfo {author}
  {\bibfnamefont {T.}~\bibnamefont {Kuroda}}, \bibinfo {author} {\bibfnamefont
  {T.}~\bibnamefont {Mano}}, \bibinfo {author} {\bibfnamefont {D.}~\bibnamefont
  {Paget}}, \bibinfo {author} {\bibfnamefont {O.}~\bibnamefont {Krebs}},
  \bibinfo {author} {\bibfnamefont {X.}~\bibnamefont {Marie}}, \bibinfo
  {author} {\bibfnamefont {K.}~\bibnamefont {Sakoda}}, \ and\ \bibinfo {author}
  {\bibfnamefont {B.}~\bibnamefont {Urbaszek}},\ }\bibfield  {title} {\enquote
  {\bibinfo {title} {Nuclear magnetization in gallium arsenide quantum dots at
  zero magnetic field},}\ }\href {\doibase 10.1038/ncomms4268} {\bibfield
  {journal} {\bibinfo  {journal} {Nat. Commun.}\ }\textbf {\bibinfo {volume}
  {5:3268}} (\bibinfo {year} {2014}),\ 10.1038/ncomms4268}\BibitemShut
  {NoStop}%
\bibitem [{\citenamefont {Zinoni}\ \emph {et~al.}(2006)\citenamefont {Zinoni},
  \citenamefont {Alloing}, \citenamefont {Monat}, \citenamefont {Zwiller},
  \citenamefont {Li}, \citenamefont {Fiore}, \citenamefont {Lunghi~s},
  \citenamefont {Gerardino}, \citenamefont {de~Riedmatten}, \citenamefont
  {Zbinden},\ and\ \citenamefont {Gisin}}]{Zinoni_APL06}%
  \BibitemOpen
  \bibfield  {author} {\bibinfo {author} {\bibfnamefont {C.}~\bibnamefont
  {Zinoni}}, \bibinfo {author} {\bibfnamefont {B.}~\bibnamefont {Alloing}},
  \bibinfo {author} {\bibfnamefont {C.}~\bibnamefont {Monat}}, \bibinfo
  {author} {\bibfnamefont {V.}~\bibnamefont {Zwiller}}, \bibinfo {author}
  {\bibfnamefont {L.~H.}\ \bibnamefont {Li}}, \bibinfo {author} {\bibfnamefont
  {A.}~\bibnamefont {Fiore}}, \bibinfo {author} {\bibfnamefont
  {L.}~\bibnamefont {Lunghi~s}}, \bibinfo {author} {\bibfnamefont
  {A.}~\bibnamefont {Gerardino}}, \bibinfo {author} {\bibfnamefont
  {H.}~\bibnamefont {de~Riedmatten}}, \bibinfo {author} {\bibfnamefont
  {H.}~\bibnamefont {Zbinden}}, \ and\ \bibinfo {author} {\bibfnamefont
  {N.}~\bibnamefont {Gisin}},\ }\bibfield  {title} {\enquote {\bibinfo {title}
  {Time-resolved and antibunching experiments on single quantum dots at
  1300nm},}\ }\href {\doibase http://dx.doi.org/10.1063/1.2190466} {\bibfield
  {journal} {\bibinfo  {journal} {Appl. Phys. Lett.}\ }\textbf {\bibinfo
  {volume} {88}},\ \bibinfo {eid} {131102} (\bibinfo {year}
  {2006})}\BibitemShut {NoStop}%
\bibitem [{\citenamefont {Takemoto}\ \emph {et~al.}(2007)\citenamefont
  {Takemoto}, \citenamefont {Takatsu}, \citenamefont {Hirose}, \citenamefont
  {Yokoyama}, \citenamefont {Sakuma}, \citenamefont {Usuki}, \citenamefont
  {Miyazawa},\ and\ \citenamefont {Arakawa}}]{Takemoto_JAP07}%
  \BibitemOpen
  \bibfield  {author} {\bibinfo {author} {\bibfnamefont {K.}~\bibnamefont
  {Takemoto}}, \bibinfo {author} {\bibfnamefont {M.}~\bibnamefont {Takatsu}},
  \bibinfo {author} {\bibfnamefont {S.}~\bibnamefont {Hirose}}, \bibinfo
  {author} {\bibfnamefont {N.}~\bibnamefont {Yokoyama}}, \bibinfo {author}
  {\bibfnamefont {Y.}~\bibnamefont {Sakuma}}, \bibinfo {author} {\bibfnamefont
  {T.}~\bibnamefont {Usuki}}, \bibinfo {author} {\bibfnamefont
  {T.}~\bibnamefont {Miyazawa}}, \ and\ \bibinfo {author} {\bibfnamefont
  {Y.}~\bibnamefont {Arakawa}},\ }\bibfield  {title} {\enquote {\bibinfo
  {title} {An optical horn structure for single-photon source using quantum
  dots at telecommunication wavelength},}\ }\href {\doibase
  http://dx.doi.org/10.1063/1.2723177} {\bibfield  {journal} {\bibinfo
  {journal} {J. Appl. Phys.}\ }\textbf {\bibinfo {volume} {101}},\ \bibinfo
  {eid} {081720} (\bibinfo {year} {2007})}\BibitemShut {NoStop}%
\bibitem [{\citenamefont {Miska}\ \emph {et~al.}(2008)\citenamefont {Miska},
  \citenamefont {Even}, \citenamefont {Dehaese},\ and\ \citenamefont
  {Marie}}]{Miska_APL08}%
  \BibitemOpen
  \bibfield  {author} {\bibinfo {author} {\bibfnamefont {P.}~\bibnamefont
  {Miska}}, \bibinfo {author} {\bibfnamefont {J.}~\bibnamefont {Even}},
  \bibinfo {author} {\bibfnamefont {O.}~\bibnamefont {Dehaese}}, \ and\
  \bibinfo {author} {\bibfnamefont {X.}~\bibnamefont {Marie}},\ }\bibfield
  {title} {\enquote {\bibinfo {title} {Carrier relaxation dynamics in
  {InAs/InP} quantum dots},}\ }\href {\doibase
  http://dx.doi.org/10.1063/1.2909536} {\bibfield  {journal} {\bibinfo
  {journal} {Appl. Phys. Lett.}\ }\textbf {\bibinfo {volume} {92}},\ \bibinfo
  {eid} {191103} (\bibinfo {year} {2008})}\BibitemShut {NoStop}%
\bibitem [{\citenamefont {Abbarchi}\ \emph {et~al.}(2010)\citenamefont
  {Abbarchi}, \citenamefont {Kuroda}, \citenamefont {Mano}, \citenamefont
  {Sakoda}, \citenamefont {Mastrandrea}, \citenamefont {Vinattieri},
  \citenamefont {Gurioli},\ and\ \citenamefont {Tsuchiya}}]{Abbarchi2010}%
  \BibitemOpen
  \bibfield  {author} {\bibinfo {author} {\bibfnamefont {M.}~\bibnamefont
  {Abbarchi}}, \bibinfo {author} {\bibfnamefont {T.}~\bibnamefont {Kuroda}},
  \bibinfo {author} {\bibfnamefont {T.}~\bibnamefont {Mano}}, \bibinfo {author}
  {\bibfnamefont {K.}~\bibnamefont {Sakoda}}, \bibinfo {author} {\bibfnamefont
  {C.~A.}\ \bibnamefont {Mastrandrea}}, \bibinfo {author} {\bibfnamefont
  {A.}~\bibnamefont {Vinattieri}}, \bibinfo {author} {\bibfnamefont
  {M.}~\bibnamefont {Gurioli}}, \ and\ \bibinfo {author} {\bibfnamefont
  {T.}~\bibnamefont {Tsuchiya}},\ }\bibfield  {title} {\enquote {\bibinfo
  {title} {Energy renormalization of exciton complexes in {GaAs} quantum
  dots},}\ }\href {\doibase 10.1103/PhysRevB.82.201301} {\bibfield  {journal}
  {\bibinfo  {journal} {Phys. Rev. B}\ }\textbf {\bibinfo {volume} {82}},\
  \bibinfo {pages} {201301(R)} (\bibinfo {year} {2010})}\BibitemShut {NoStop}%
\bibitem [{\citenamefont {Mlinar}\ and\ \citenamefont
  {Zunger}(2009)}]{Mlinar_PRB09}%
  \BibitemOpen
  \bibfield  {author} {\bibinfo {author} {\bibfnamefont {V.}~\bibnamefont
  {Mlinar}}\ and\ \bibinfo {author} {\bibfnamefont {A.}~\bibnamefont
  {Zunger}},\ }\bibfield  {title} {\enquote {\bibinfo {title} {Spectral
  barcoding of quantum dots: Deciphering structural motifs from the excitonic
  spectra},}\ }\href {\doibase 10.1103/PhysRevB.80.035328} {\bibfield
  {journal} {\bibinfo  {journal} {Phys. Rev. B}\ }\textbf {\bibinfo {volume}
  {80}},\ \bibinfo {pages} {035328} (\bibinfo {year} {2009})}\BibitemShut
  {NoStop}%
\bibitem [{\citenamefont {Kowalik}\ \emph {et~al.}(2007)\citenamefont
  {Kowalik}, \citenamefont {Krebs}, \citenamefont {Lema{\^i}tre}, \citenamefont
  {Eble}, \citenamefont {Kudelski}, \citenamefont {Voisin}, \citenamefont
  {Seidl},\ and\ \citenamefont {Gaj}}]{Kowalik_APL07}%
  \BibitemOpen
  \bibfield  {author} {\bibinfo {author} {\bibfnamefont {K.}~\bibnamefont
  {Kowalik}}, \bibinfo {author} {\bibfnamefont {O.}~\bibnamefont {Krebs}},
  \bibinfo {author} {\bibfnamefont {A.}~\bibnamefont {Lema{\^i}tre}}, \bibinfo
  {author} {\bibfnamefont {B.}~\bibnamefont {Eble}}, \bibinfo {author}
  {\bibfnamefont {A.}~\bibnamefont {Kudelski}}, \bibinfo {author}
  {\bibfnamefont {P.}~\bibnamefont {Voisin}}, \bibinfo {author} {\bibfnamefont
  {S.}~\bibnamefont {Seidl}}, \ and\ \bibinfo {author} {\bibfnamefont {J.~A.}\
  \bibnamefont {Gaj}},\ }\bibfield  {title} {\enquote {\bibinfo {title}
  {Monitoring electrically driven cancellation of exciton fine structure in a
  semiconductor quantum dot by optical orientation},}\ }\href {\doibase
  http://dx.doi.org/10.1063/1.2805025} {\bibfield  {journal} {\bibinfo
  {journal} {Appl. Phys. Lett.}\ }\textbf {\bibinfo {volume} {91}},\ \bibinfo
  {eid} {183104} (\bibinfo {year} {2007})}\BibitemShut {NoStop}%
\bibitem [{\citenamefont {Belhadj}\ \emph {et~al.}(2009)\citenamefont
  {Belhadj}, \citenamefont {Simon}, \citenamefont {Amand}, \citenamefont
  {Renucci}, \citenamefont {Chatel}, \citenamefont {Krebs}, \citenamefont
  {Lema\^itre}, \citenamefont {Voisin}, \citenamefont {Marie},\ and\
  \citenamefont {Urbaszek}}]{Belhadj_PRL09}%
  \BibitemOpen
  \bibfield  {author} {\bibinfo {author} {\bibfnamefont {T.}~\bibnamefont
  {Belhadj}}, \bibinfo {author} {\bibfnamefont {C.-M.}\ \bibnamefont {Simon}},
  \bibinfo {author} {\bibfnamefont {T.}~\bibnamefont {Amand}}, \bibinfo
  {author} {\bibfnamefont {P.}~\bibnamefont {Renucci}}, \bibinfo {author}
  {\bibfnamefont {B.}~\bibnamefont {Chatel}}, \bibinfo {author} {\bibfnamefont
  {O.}~\bibnamefont {Krebs}}, \bibinfo {author} {\bibfnamefont
  {A.}~\bibnamefont {Lema\^itre}}, \bibinfo {author} {\bibfnamefont
  {P.}~\bibnamefont {Voisin}}, \bibinfo {author} {\bibfnamefont
  {X.}~\bibnamefont {Marie}}, \ and\ \bibinfo {author} {\bibfnamefont
  {B.}~\bibnamefont {Urbaszek}},\ }\bibfield  {title} {\enquote {\bibinfo
  {title} {Controlling the polarization eigenstate of a quantum dot exciton
  with light},}\ }\href {\doibase 10.1103/PhysRevLett.103.086601} {\bibfield
  {journal} {\bibinfo  {journal} {Phys. Rev. Lett.}\ }\textbf {\bibinfo
  {volume} {103}},\ \bibinfo {pages} {086601} (\bibinfo {year}
  {2009})}\BibitemShut {NoStop}%
\bibitem [{\citenamefont {Cortez}\ \emph {et~al.}(2002)\citenamefont {Cortez},
  \citenamefont {Krebs}, \citenamefont {Laurent}, \citenamefont {Senes},
  \citenamefont {Marie}, \citenamefont {Voisin}, \citenamefont {Ferreira},
  \citenamefont {Bastard}, \citenamefont {G\'erard},\ and\ \citenamefont
  {Amand}}]{Cortez_PRL02}%
  \BibitemOpen
  \bibfield  {author} {\bibinfo {author} {\bibfnamefont {S.}~\bibnamefont
  {Cortez}}, \bibinfo {author} {\bibfnamefont {O.}~\bibnamefont {Krebs}},
  \bibinfo {author} {\bibfnamefont {S.}~\bibnamefont {Laurent}}, \bibinfo
  {author} {\bibfnamefont {M.}~\bibnamefont {Senes}}, \bibinfo {author}
  {\bibfnamefont {X.}~\bibnamefont {Marie}}, \bibinfo {author} {\bibfnamefont
  {P.}~\bibnamefont {Voisin}}, \bibinfo {author} {\bibfnamefont
  {R.}~\bibnamefont {Ferreira}}, \bibinfo {author} {\bibfnamefont
  {G.}~\bibnamefont {Bastard}}, \bibinfo {author} {\bibfnamefont {J.-M.}\
  \bibnamefont {G\'erard}}, \ and\ \bibinfo {author} {\bibfnamefont
  {T.}~\bibnamefont {Amand}},\ }\bibfield  {title} {\enquote {\bibinfo {title}
  {Optically driven spin memory in n-doped {InAs-GaAs} quantum dots},}\ }\href
  {\doibase 10.1103/PhysRevLett.89.207401} {\bibfield  {journal} {\bibinfo
  {journal} {Phys. Rev. Lett.}\ }\textbf {\bibinfo {volume} {89}},\ \bibinfo
  {pages} {207401} (\bibinfo {year} {2002})}\BibitemShut {NoStop}%
\bibitem [{\citenamefont {Bracker}\ \emph {et~al.}(2005)\citenamefont
  {Bracker}, \citenamefont {Stinaff}, \citenamefont {Gammon}, \citenamefont
  {Ware}, \citenamefont {Tischler}, \citenamefont {Shabaev}, \citenamefont
  {Efros}, \citenamefont {Park}, \citenamefont {Gershoni}, \citenamefont
  {Korenev},\ and\ \citenamefont {Merkulov}}]{Bracker_PRL05}%
  \BibitemOpen
  \bibfield  {author} {\bibinfo {author} {\bibfnamefont {A.~S.}\ \bibnamefont
  {Bracker}}, \bibinfo {author} {\bibfnamefont {E.~A.}\ \bibnamefont
  {Stinaff}}, \bibinfo {author} {\bibfnamefont {D.}~\bibnamefont {Gammon}},
  \bibinfo {author} {\bibfnamefont {M.~E.}\ \bibnamefont {Ware}}, \bibinfo
  {author} {\bibfnamefont {J.~G.}\ \bibnamefont {Tischler}}, \bibinfo {author}
  {\bibfnamefont {A.}~\bibnamefont {Shabaev}}, \bibinfo {author} {\bibfnamefont
  {A.~L.}\ \bibnamefont {Efros}}, \bibinfo {author} {\bibfnamefont
  {D.}~\bibnamefont {Park}}, \bibinfo {author} {\bibfnamefont {D.}~\bibnamefont
  {Gershoni}}, \bibinfo {author} {\bibfnamefont {V.~L.}\ \bibnamefont
  {Korenev}}, \ and\ \bibinfo {author} {\bibfnamefont {I.~A.}\ \bibnamefont
  {Merkulov}},\ }\bibfield  {title} {\enquote {\bibinfo {title} {Optical
  pumping of the electronic and nuclear spin of single charge-tunable quantum
  dots},}\ }\href {\doibase 10.1103/PhysRevLett.94.047402} {\bibfield
  {journal} {\bibinfo  {journal} {Phys. Rev. Lett.}\ }\textbf {\bibinfo
  {volume} {94}},\ \bibinfo {pages} {047402} (\bibinfo {year}
  {2005})}\BibitemShut {NoStop}%
\bibitem [{\citenamefont {Bester}, \citenamefont {Nair},\ and\ \citenamefont
  {Zunger}(2003)}]{Bester_PRB03}%
  \BibitemOpen
  \bibfield  {author} {\bibinfo {author} {\bibfnamefont {G.}~\bibnamefont
  {Bester}}, \bibinfo {author} {\bibfnamefont {S.}~\bibnamefont {Nair}}, \ and\
  \bibinfo {author} {\bibfnamefont {A.}~\bibnamefont {Zunger}},\ }\bibfield
  {title} {\enquote {\bibinfo {title} {Pseudopotential calculation of the
  excitonic fine structure of million-atom self-assembled
  {In}$_{1-x}${Ga}$_x${As/GaAs} quantum dots},}\ }\href {\doibase
  10.1103/PhysRevB.67.161306} {\bibfield  {journal} {\bibinfo  {journal} {Phys.
  Rev. B}\ }\textbf {\bibinfo {volume} {67}},\ \bibinfo {pages} {161306}
  (\bibinfo {year} {2003})}\BibitemShut {NoStop}%
\bibitem [{\citenamefont {He}\ \emph {et~al.}(2008)\citenamefont {He},
  \citenamefont {Gong}, \citenamefont {Li}, \citenamefont {Guo},\ and\
  \citenamefont {Zunger}}]{He_PRL08}%
  \BibitemOpen
  \bibfield  {author} {\bibinfo {author} {\bibfnamefont {L.}~\bibnamefont
  {He}}, \bibinfo {author} {\bibfnamefont {M.}~\bibnamefont {Gong}}, \bibinfo
  {author} {\bibfnamefont {C.-F.}\ \bibnamefont {Li}}, \bibinfo {author}
  {\bibfnamefont {G.-C.}\ \bibnamefont {Guo}}, \ and\ \bibinfo {author}
  {\bibfnamefont {A.}~\bibnamefont {Zunger}},\ }\bibfield  {title} {\enquote
  {\bibinfo {title} {Highly reduced fine-structure splitting in {InAs/InP}
  quantum dots offering an efficient on-demand entangled 1.55-$\mu${}m photon
  emitter},}\ }\href {\doibase 10.1103/PhysRevLett.101.157405} {\bibfield
  {journal} {\bibinfo  {journal} {Phys. Rev. Lett.}\ }\textbf {\bibinfo
  {volume} {101}},\ \bibinfo {pages} {157405} (\bibinfo {year}
  {2008})}\BibitemShut {NoStop}%
\bibitem [{\citenamefont {Goldmann}\ \emph {et~al.}(2013)\citenamefont
  {Goldmann}, \citenamefont {Barthel}, \citenamefont {Florian}, \citenamefont
  {Schuh},\ and\ \citenamefont {Jahnke}}]{Goldmann_APL13}%
  \BibitemOpen
  \bibfield  {author} {\bibinfo {author} {\bibfnamefont {E.}~\bibnamefont
  {Goldmann}}, \bibinfo {author} {\bibfnamefont {S.}~\bibnamefont {Barthel}},
  \bibinfo {author} {\bibfnamefont {M.}~\bibnamefont {Florian}}, \bibinfo
  {author} {\bibfnamefont {K.}~\bibnamefont {Schuh}}, \ and\ \bibinfo {author}
  {\bibfnamefont {F.}~\bibnamefont {Jahnke}},\ }\bibfield  {title} {\enquote
  {\bibinfo {title} {Excitonic fine-structure splitting in telecom-wavelength
  {InAs/GaAs} quantum dots: Statistical distribution and height-dependence},}\
  }\href {\doibase http://dx.doi.org/10.1063/1.4833027} {\bibfield  {journal}
  {\bibinfo  {journal} {Appl. Phys. Lett.}\ }\textbf {\bibinfo {volume}
  {103}},\ \bibinfo {eid} {242102} (\bibinfo {year} {2013})}\BibitemShut
  {NoStop}%
\end{thebibliography}%

\end{document}